\documentclass[a4,11pt,fleqn]{article}
\usepackage{latexsym}
\usepackage{amsmath}
\usepackage{amssymb}
\usepackage[dvips]{graphicx}
\author{  \quad \\ Luigi  Sertorio $^{\dag}$  \and   \quad \\ Giovanna  Tinetti 
$^{\ddag}$ }
\title{\textbf{{\LARGE Constraints in the coupling Star-Life  }}}
\date{  {\footnotesize \emph{Dipartimento di Fisica Teorica -
Universit\`a di Torino. \\ Istituto Nazionale di Fisica Nucleare -
Sezione di Torino}} \\ \quad \\ {\footnotesize DFTT 27/2001} \\ \quad \\
{\footnotesize $^{\dag}$ sertorio@to.infn.it \qquad \qquad  
 $^{\ddag}$  tinetti@to.infn.it } }
\begin{document}
\maketitle
\renewcommand{\theequation}{\thesection.\arabic{equation}}
\renewcommand{\thefigure}{\thesection.\arabic{figure}}
\newcommand{\ud}{\mathrm{d}}
\newcommand{\de}{\partial}
\section*{Abstract}
\begin{tabular}{|p{14. cm}|}
\hline
{\small
If life is sustained by a process of photosynthesis, not necessarily the same existing on Earth, the surface temperature of the star 
and the orbit of the host planet  cannot be whatsoever.
In fact the global life cycle, no matter how complicated, must contain in general an upper photochemical branch and a lower dark branch, characterized by a higher and a lower temperature (\S2).
These two temperatures are star-orbit related. The velocity along the cycle or, in other words, the power of the life machine,
 depends in general on several other parameters. First of all the Gibbs photon availability, which is a star-orbit parameter and is the input for the upper branch (\S3). Then follows the energy cascade that develops along the organic web with a large number of interactions and typical times that must match the typical times  generated by the combination of spin value and orientation, eccentricity and precession (\S4). 
Finally, the capacity of the web to keep the global life
cycle running along the life span of the star, comes from some inner form of self-endurance and self-balance. The property of not being transient could be the right way of introducing the concept of intelligent life (\S1).

\par }
    \\
\hline
\end{tabular}   \newpage
\section{Life from inside}
\setcounter{equation}{0}
\setcounter{figure}{0}
We need to face a fundamental distinction, namely among deterministic and finalistic systems. 

Deterministic systems are described by equations of motion, either known explicitly or supposed to exist, which come from a minimum principle related to a Lagrangian.
This is the domain of physics. 

Finalistic systems are those in which what is minimized is another functional that can never be reduced  to a Lagrangian type, because the object to be minimized contains ``external inputs''.  This is a formal way of introducing the concept of finalism and can be helpful, although not necessarily the profoundest. 

A finite size finalistic system is realized by a standing chemical structure, occupying a certain domain in space. 
Such standing finite structure requires a power density. This object can be called organism, including with this word a large variety of structural complexities and control abilities. Going on in the process of considering complex structures and related complex control functions, we cross a borderline and talk of living organisms.

Finally we wish to touch only marginally the concept of intelligent life. This concept is particularly ambiguous, because the only way we know to communicate acquired knowledge is a forward process both for an individual and for a global community. Even the most abstract way of thinking, mathematics, develops in time. Therefore the question is: intelligent at what \emph{time},
 rather than intelligent up to a certain \emph{score}.
It follows that for a given realization of life, for instance that on Earth, or another,  rather than abstract intelligence,
we should  talk of history of intelligence assessed in a particular time interval of the presumed course of development. 
This also implies typical times. For instance the human terrestrial civilization is understandable for, say, few thousand years.
This may be a very long time interval for another form of life on a different  planet, or extremely short, almost not noticeable, for another. In all cases, it is the global cycle, and its history, that contains the maximum amount of intelligence that was given to that particular planet.

In conclusion, the discussion of intelligent life referred to another form of life existing elsewhere, remains a rather uncertain issue.

\section{Life from outside}
\setcounter{equation}{0}
\setcounter{figure}{0}
The analytic knowledge of a living organism assumes, as a methodology, the consideration of a simple component, or elementary organism, then several components constituting a higher organism, and so on. Finally animals, and man, supposed to be the best, or highest 
 organism. Let us go on, a step further, and consider the whole of the living system, however complicated. This is what is seen from outside.

The complex operation of this ``global organism'' is peculiar. In fact the power input is given uniquely by the incoming photons and the power output is again only by photons, those leaving the planet. This is the way in which global life is organized,
 if it is photon life.
So while every living organism has an ambient, an ecosystem around it, the global organism is that one having for ecosystem the entire Universe, appearing via the star nearby and the cosmological blackbody background radiation.

We have considered in a preceding paper \cite{ref:ava} the case of a generic life based on a process of mining energy and material, and based on a process of extracting the thermal availability embedded in a thermal disequilibrium flow; this is the case of a planet without star, which slowly cools off and therefore contains thermal gradients.
We belong to photon induced life. We wish to consider this case in general and not uniquely in the terrestrial realization.

Photon life is a system open to energy in the form of in and out radiation fluxes. Concerning the material transit across the living web, we have the important concept of asymptotic constituents.
Within the global organism a large number of complicated assemblies of molecules, organisms, are formed, stand for a while,
 and decay.
In doing so, they communicate material and energy to each other. This is the global metabolism. Nevertheless, this global metabolism
rests on the circulation of asymptotic molecules; those, among the released, which are capable of being photo-synthesized again.
For instance, in the particular case of the terrestrial life, the asymptotic molecules are H$_{2}$O and CO$_{2}$.
There is no global mining and correspondingly accumulation of products unable to return to the mine.

The asymptotic molecules come out of the web at a certain rate, expressed in mass per second, move along the planet in either liquid or gas phase, and then return again as components in the process of photosynthesis to restart the construction of complex structures,
 which keep feeding the web, and so on forever, in principle. The global rate of photosynthesis must be equal to the global rate of dark release \cite{ref:sup}, if the global circulation 
is stationary. It is in this sense that the molecules H$_{2}$O and CO$_{2}$ are not only participating in the dynamics, 
as any other component of the living web, where nothing is fixed, but in addition are asymptotic. If the global life were terminated, all the molecules participating to the standing global structure would be released on the planet and would settle down to the final distribution, namely that corresponding to the minimum of the Gibbs potential, at those values of $T$, and $P$, which are pertinent to the planet.

\begin{figure}[ptbh]
\begin{center}
 \includegraphics[width=7 cm]{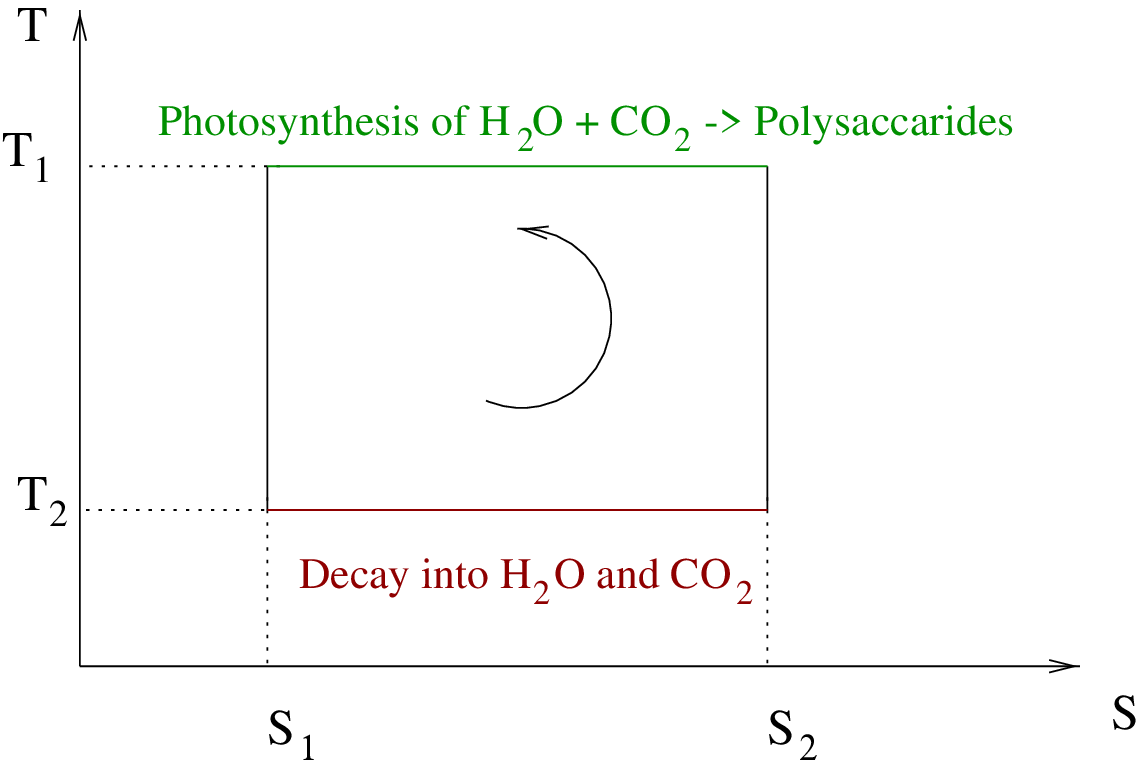}  
\caption{ \emph{ {\footnotesize 
 }}}  \label{fig:g1}
\end{center}
\end{figure}

From the viewpoint of energy and entropy, the global web corresponds to this cycle in the diagram temperature-entropy 
(fig. \ref{fig:g1}).  \\
 \\
$T_{1}$ is the Planck spectrum temperature of the photons activating the chlorophyll, $T_{2}$ is the dark temperature of the final decay which ultimately must reproduce H$_{2}$O and CO$_{2}$ in the right amount that is required to restart the cycle on the upper branch at $T_{1}$. 

\begin{figure}[ptbh]
\begin{center}
\includegraphics[width=14 cm]{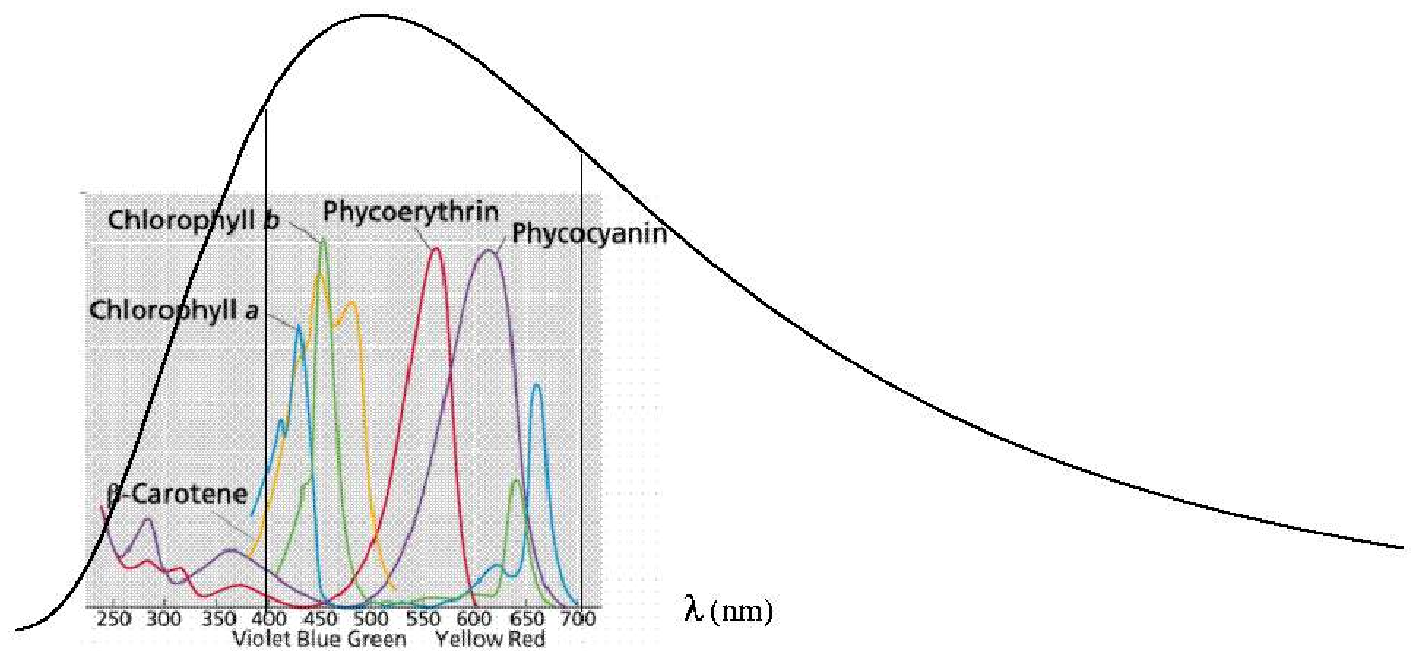}  
\caption{ \emph{ {\footnotesize 
 }}}  \label{fig:g2}
\end{center}
\end{figure}

We show in fig. \ref{fig:g2} the window of frequencies able to operate the photosynthesis in the case of the terrestrial carbon life.
For another triplet $\tilde{H}, \tilde{C}, \tilde{O}$ we may expect another $T_{s}$ and another wavelength window. \\

The energy balance implies $E_{1}$ incoming with the photon flux originated by the star, $E_{2}$ outgoing with the photon flux leaving the planet, which acts as the thermal bath at $T_{2}$. The global cycle is stationary if
\begin{equation*}
E_{1} = E_{2} = E
\end{equation*}
For the entropy $S$ we have
\begin{equation*} 
S_{1} = S^{in} = \frac{E}{T_{1}}
\end{equation*}
\begin{equation*} 
S_{2} = S^{out} = \frac{E}{T_{2}}
\end{equation*}
\begin{equation*} 
 S^{out} -  S^{in} = \Delta S = E \, \left ( \frac{1}{T_{2}} - \frac{1}{T_{1}} \right ) = \frac{E}{T_{2}} \,
\left ( 1 - \frac{T_{2}}{T_{1}} \right )
\end{equation*}
Let us consider now the concept of availability.
The Gibbs availability \cite{ref:ava} corresponding to the standing structure which  performs the cycle is 
\begin{equation*} 
\Delta W = T_{2} \, \Delta S =   E \,
\left ( 1 - \frac{T_{2}}{T_{1}} \right )
\end{equation*}
Notice that we expect $T_{2} \ll T_{1}$ in general.
At $T_{2}$ the asymptotic molecules  perform the decay reactions of the lower branch, or the dark branch.
To fix the ideas, for the chemical life on Earth based on C, O, H, N, we have
$T_{1} = T_{\odot} \sim 5800 \, \text{K}$,
$T_{2} \sim 300 \, \text{K}$ and
\begin{equation*} 
\Delta W \sim E
\end{equation*}
Now, much more interesting than the above equilibrium thermodynamic description of the cycle, would be the understanding of the
non-equilibrium thermodynamic description, that is the time rate of the various processes (see \S 3).
From such an analysis, we could formulate the concept of power of the global ecosystem cycle. 
In fact the photons come as a flux, energy per unit time, per unit surface.
They trigger all the dynamical interactions in the living web. These interactions have a large number of typical dynamical times.
Consider for example the time pertinent to an electronic molecular quantum level excitations, $\tau \sim 10^{-15} \, \text{s}$;
and consider for instance, on the other extreme, the life time of a red tree
\begin{equation*} 
\tau \sim 2000 \, \text{years} \sim 6 \cdot 10^{10} \, \text{s}
\end{equation*}
Between these two extremes, so many other typical times are excited and all of them must function combining with each other to produce
the actual power balance
\begin{equation*} 
\Phi^{in}_{bio} = \Phi^{out}_{bio}
\end{equation*}
where $\Phi^{in}_{bio}$ is the portion of the radiation flux coming from the Sun  absorbed by the living ecosystem, and 
$\Phi^{out}_{bio}$ is the contribution of the ecosystem to the radiation flux emitted by the Earth. 

Another form of life may function with asymptotic components $\tilde{H}, \tilde{C}, \tilde{O}$, which are synthesized into another structure, a polymer $\left \{ \tilde{H}, \tilde{C}, \tilde{O} \right \}$, requiring a spectral temperature $\tilde{T}_{s}$ and a dark temperature
$\tilde{T}$.
Different sets of typical times will be involved. But not whatsoever. We know that the laws of quantum mechanics must be obeyed,
therefore we may think of a  different chemical life but within the identical chemistry. 

The intermediary between a form of life and its star is the ``protagonist planet'' and its orbit.
The typical times generated by the orbital parameters on one side, and  the typical times of the unknown form of life,
on the other side, must combine together. This is what we mean by ``constraint'' in the present paper.

\section{The two temperatures and the year}
\setcounter{equation}{0}
\setcounter{figure}{0}
Consider now a planet that contains the appropriate set of molecules which can organize themselves in complex chemical 
structures that   eventually evolve into life.
Among these molecules there must be the subset of asymptotic molecules, not necessarily H$_{2}$O and CO$_{2}$.
The minimum conditions for the existence of life
are given by the existence of the photon temperature $T_{s}$ and the dark temperature $T$. 

\begin{figure}[ptbh]
\begin{center}
 \includegraphics[width=7 cm]{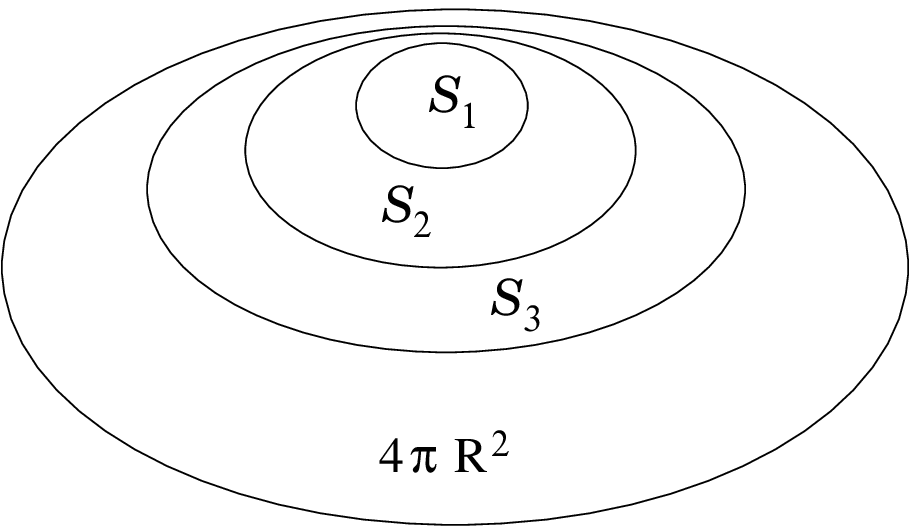}  
\caption{ \emph{ {\footnotesize 
 }}}  \label{fig:b1}
\end{center}
\end{figure}

Notice that the life cycle dark branch implies not only a dark $T$ but
also a dark pressure $P$.
 The pressure $P$ is  a parameter that comes from the history of the  formation of the planet itself. This history
 determines if there is an abundant atmosphere or not, and consequently
the value of $P$.
What is important is the constraint that the asymptotic molecules be in fluid state to ensure the necessary supply-flow of matter 
required by the incoming photons to perform their synthesis job.
In fact the photosynthesis is a high energy density process and is fast; the corresponding dark decay is a low energy density process
and is slow. To close the cycle a large surface is necessary in order to sustain the transport of matter towards the synthesis centers, hence the fluid motions.
Roughly speaking we can think of this balance:
\begin{equation} \label{eq:b1}
\Phi_{bio}^{in} = 
\sigma \, T_{s}^{4} \,  \mathcal{S}_{1}
\end{equation}
where $\mathcal{S}_{1}$ is the total area of the photosynthesis centers. $\mathcal{S}_{1}$ can be thought, in the language
of elementary particle physics, as the total cross section of the reaction
\begin{equation*}
\gamma + \text{center} \, \to \, \text{inner operation on H}_{2}\text{O, CO}_{2} \, \to \, \text{polisaccharide} 
\end{equation*}
At the lower branch the organism decays into the asymptotic molecules that thermalize at $T$ and uniquely release the corresponding heat of the exothermic reaction into radiation with density of power $\sigma \, T^{4}$.
The total outgoing flux is
\begin{equation} \label{eq:b2}
\Phi_{bio}^{out} = 
\sigma \, T^{4} \,  \mathcal{S}_{3}
\end{equation}
What is $\mathcal{S}_{3}$? Consider fig. \ref{fig:b1}

$\mathcal{S}_{1}$ is the area of the photosynthetic center \\
$\mathcal{S}_{2}$ is the area occupied by living organism \\
$\mathcal{S}_{3}$ is the effective area needed by the asymptotic molecules to release their heat of reaction at temperature $T$  \\
$4 \, \pi \, R^{2}$ is the total emitting surface of the planet. The closure of the cycle implies
\begin{equation*} 
\Phi_{bio}^{in} = \Phi_{bio}^{out}
\end{equation*}
namely 
\begin{equation*} 
\sigma \, T_{s}^{4} \,  \mathcal{S}_{1} = \sigma \, T^{4} \,  \mathcal{S}_{3}
\end{equation*}
and this equation implies
\begin{equation} \label{eq:b4}
\beta = \frac{\mathcal{S}_{1}}{\mathcal{S}_{3}} = \frac{T^{4}}{T_{s}^{4}} 
\end{equation}
From appendix B, we have
\begin{equation*} 
 \frac{T^{4}}{T_{s}^{4}} = \frac{R_{s}^{2}}{d^{2}}
\end{equation*}
therefore $\beta$ is dictated by astronomy. We expect
\begin{equation*} 
\beta = \frac{\mathcal{S}_{1}}{\mathcal{S}_{3}} 
\end{equation*}
to be a rather small number. For instance in the particular case of the Earth we have $T_{s} = 6000 \, \text{K}$,
$T = 300 \, \text{K}$ and therefore
\begin{equation} \label{eq:b5}
\beta \sim 0.6 \cdot 10^{-5} 
\end{equation}
An interesting question is: how large is $\mathcal{S}_{3}$ with respect to $4 \, \pi \, R^{2}$?
Reference \cite{ref:mar} gives for
\begin{equation*}
\Phi_{bio}^{in} = 10^{15} \, \text{W}
\end{equation*}
so that we can write 
 \begin{equation}  \label{eq:str}
 \sigma \, T^{4} \,  \mathcal{S}_{3}  = 10^{15} \, \text{W}
\end{equation}
On the other hand we know that 
\begin{equation} \label{eq:str1}
4 \, \pi \, R^{2} \, \sigma \, T^{4} \,   = 1.2 \cdot 10^{17} \, \text{W}
\end{equation}
Comparing (\ref{eq:str}) and (\ref{eq:str1}) we get
\begin{equation} \label{eq:str2}
\mathcal{S}_{3}  \sim 10^{-2} \cdot 4 \, \pi \, R^{2}    
\end{equation}

The evaluation (\ref{eq:b5}) indicates quite clearly that we need fluid transport of the asymptotic molecules;
 hence it makes sense to talk
not only of a temperature  $T$, but of an appropriate pressure $P$.

How can we predict the temperature distribution over the protagonist planet?
It is almost impossible to describe in general the fluid motions existing over an abstract planet. 
The best knowledge we have about the dissipative structures, comes from the simple and well studied case of the B\'enard cells.
What we know qualitatively, in general, is that the presence of the fluid circulation increases the heat transport with respect to the transport described by the pure rigid planet, that is the heat transport described by the  Fourier equation \cite{ref:nc}.

In conclusion we study a model planet which is a rigid black sphere. There are two motivations for this study.
\begin{description}
\item[-] The thermodynamics of a rigid surface is governed by the Fourier equation, which is linear, and this property allows a rather satisfactory analytic study of the temperature field. 
\item[-] The surface temperature distribution given by the rigid model is the most reasonable input for the actual, unknown, atmospheric dynamics and corresponding heat circulation.
\end{description}
We can say that the black planet plays the role of the lower, warm, plate of B\'enard.

\emph{Note: In the particular case of the Earth, the rigid black planet
 approximation is not used as an input for the understanding of the atmospheric circulation. This is because the Earth has a known geography, and we are sitting here to observe it; this knowledge is a 
much more detailed input for the study of the fluid circulation, of course. But this is the approach from \underline{inside} and
\underline{now}. Different geography, different circulation. It is the historical anthropomorphic viewpoint.
And this viewpoint is inadequate for our purposes.
For the abstract planet which is the protagonist of the present study the black sphere model is the only model we can think of. }

We dedicate \S 4 to comments on the atmosphere.

After these preliminaries, we consider a star with surface temperature $T_{s}$ and a planet rotating around it in a closed elliptic orbit.
Rigorously speaking, a two body planetary orbit is an idealization never realized in nature on the long time scales. Nevertheless
 this assumption is very attractive, because it allows us to push forward some general analytical results, so we adopt it.

\begin{figure}[ptbh]
\begin{center}
 \includegraphics[width=8 cm]{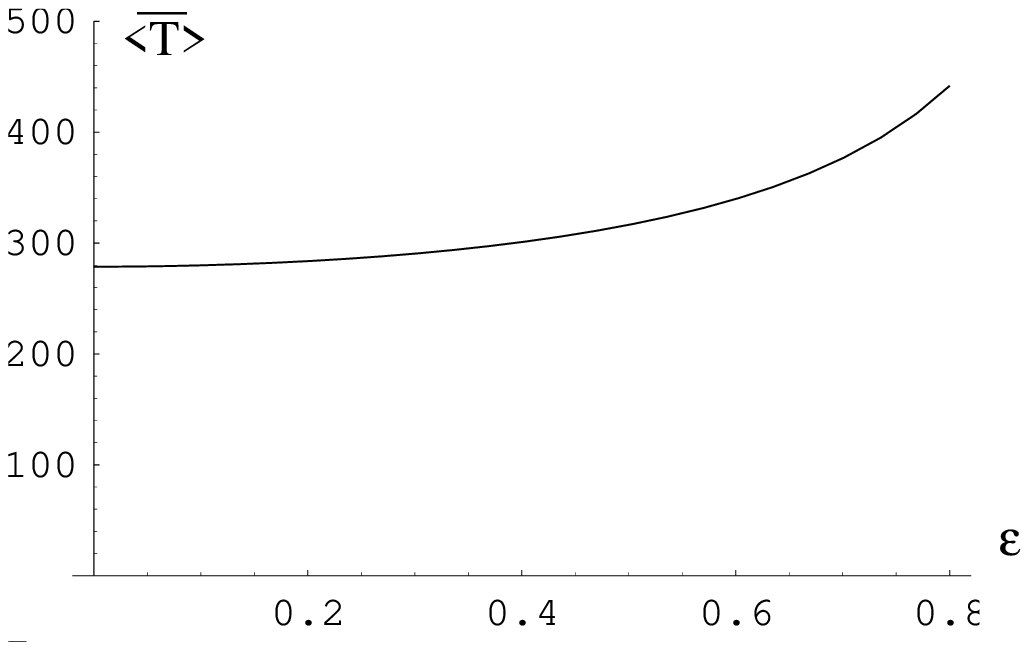}  
\caption{ \emph{ {\footnotesize 
 }}}  \label{fig:bb1}
\end{center}
\end{figure}

The planet at position $\psi$ on the ellipse receives a total radiation flux (see appendixes A, B)
\begin{equation} \label{eq:g1}
\Phi^{in}_{tot} =
\sigma \, T_{s}^{4} \, \frac{R_{S}^{2}}{d^{2} (\psi)} \,\pi \, R^{2}
\end{equation}
The photons come with Planck temperature $T_{s}$. The function $d^{2} (\psi)$ mod $2 \, \pi$ is known.
Let us assume that the incoming and outgoing fluxes are equal for any $\psi$
\begin{equation*} 
\sigma \, T_{s}^{4} \, \frac{R_{S}^{2}}{d^{2} (\psi)} \,\pi \, R^{2} = \sigma \, \overline{T^{4}} (\psi) \, 4 \, \,\pi \, R^{2}
\end{equation*}
Notice this is a power balance not a linear relationship between temperatures. From this equation we can define a
 ``radiation average'' 
\begin{equation} \label{eq:g2}
<T (\psi)> = \left [ \overline{T^{4}} (\psi)  \right ]^{\frac{1}{4}}   = \frac{1}{\sqrt{2}} \,
 \, \sqrt{\frac{R_{S}}{d (\psi)}} \, T_{s} = \frac{T_{s}}{\sqrt{2}} \,
 \sqrt{\frac{R_{S}}{a \, (1 - \epsilon^{2})}}  \, 
\sqrt{1 - \epsilon \, \sin \psi} 
\end{equation}
The amplitude of the oscillation of $<T (\psi)>$ during the year is 
\begin{equation*} 
\Delta <T> \equiv <T> (1^{st} solstice) - <T> (2^{nd} solstice) 
\end{equation*}
With our choice of the polar angle $\psi$ we have: \\
$\psi (1^{st} solstice) = - \frac{\pi}{2}$ and 
$\psi (2^{nd} solstice)  = \frac{\pi}{2}$, so we get from (\ref{eq:g2})
\begin{equation*} 
\Delta <T>  =
\frac{1}{\sqrt{2}} \,
 \, \sqrt{\frac{R_{S}}{a}} \, T_{s} \, \left [ \frac{1}{\sqrt{1 - \epsilon}} - \frac{1}{\sqrt{1 + \epsilon}}  \right  ]
\end{equation*}
Clearly this amplitude is zero for $\epsilon = 0$. \\
The annual average of $<T (\psi)>$ is 
\begin{equation} \label{eq:g3}
\begin{split}
\overline{<T>} & = \frac{1}{2 \, \pi} \, \int_{0}^{2 \, \pi} \, <T (\psi)> \, \ud \psi
= \frac{T_{s}}{\sqrt{2}} \,
 \sqrt{\frac{R_{S}}{a \, (1 - \epsilon^{2})}}  \, \frac{1}{2 \, \pi} \, \int_{0}^{2 \, \pi} \,
\sqrt{1 - \epsilon \, \sin \psi} \, \ud \psi = \\
& = \frac{1}{{\sqrt{2}}\,\pi } \, { {\sqrt{{\frac{R_{s}}{a\,\left( 1 + \epsilon  \right) }}}}\, T_{s} \,
     \left[ E \left({\frac{\pi }{4}},{\frac{-2\,\epsilon }{1 - \epsilon }} \right ) + 
       E \left ({\frac{3\,\pi }{4}},{\frac{-2\,\epsilon }{1 - \epsilon }} \right ) \right] } \\
\end{split}
\end{equation}
$E (\phi, m)$ are elliptic integrals of the second kind known in the literature \cite{ref:abr}. The dependence on $\epsilon$
of  $\overline{<T>}$ is obtained using the tabulation of      $E (\phi, m)$, (fig. \ref{fig:bb1}) \\
Finally we consider the ratio
\begin{equation*} 
\frac{\overline{<T>} (\epsilon)}{\overline{<T>} (0)} =  
{\frac{1}{\pi } \, {\sqrt{{\frac{1}{\left( 1 + \epsilon  \right) }}}}\,
    \left[ E \left({\frac{\pi }{4}},{\frac{-2\,\epsilon }{1 - \epsilon }} \right ) + 
       E \left ({\frac{3\,\pi }{4}},{\frac{-2\,\epsilon }{1 - \epsilon }} \right ) \right]    } \quad _{\overrightarrow{\epsilon \to 0}} \quad 1
\end{equation*}
This formula comes from the observation that $E (\phi, 0) = \phi $.

Now, the life processes on the planet depend on the time $t$ not on the angle $\psi$ (appendix A, eq. (\ref{eq:s23})).
Our ellipse is characterized by $a$ and $\epsilon$. 
We show in appendix A that, given $a$ and the masses of the star $M_{s}$ and the planet $m_{p}$, it follows the annual period $\tau_{y}$: 
\begin{equation} \label{eq:g4}
\tau_{y} 
 = 2 \, \pi \, \sqrt{\frac{a^{3}}{\gamma \, (M_{S} + m_{p})}}
\end{equation}
At this point we observe that our planet cannot be of the Jovian type (mostly hydrogen). Consequently we neglect $m_{p}$
in (\ref{eq:g4}) so that $\tau_{y}$ depends only on $a$ and $M_{s}$.
But we know that for stars in the Main Sequence the mass $M_{s}$ and the temperature $T_{s}$ are related. Consequently
if the hypothetical photosynthesis exists over the protagonist planet, that particular form of life implies a certain $T_{s}$
and a certain $T_{b}$ ($b$ stands for best approximation of the temperature of the dark branch).
The couple ($T_{s}, \, T_{b}$) may couple only with one couple of ($M_{s}, \, a)$; therefore the lenght of the
 year follows from the knowledge of
($T_{s}, \, T_{b}$)
\begin{equation*}
\tau_{y} 
 = \tau_{y}  (T_{s}, \, T_{b})
\end{equation*}
Let us assume for simplicity that $\epsilon = 0, \, d = a$. We use for $T_{b}$ the value (\ref{eq:g2}), so $T_{b}$
is a function of $T_{s}$ and $a$, therefore given the couple $T_{s}$ and $T_{b}$ using (\ref{eq:g4}) we get $\tau_{y}$.
We show in the table below $\tau_{y}$ as a function of $T_{s}$ and $T_{b}$, $\tau_{y}$ is expressed in units of $\tau_{ye}$,
namely the terrestrial year. On the coordinate axis $T_{s}$ we add the column $t_{s}$
(life of the star) and $M_{s}$ (mass of the star) taken from \cite{ref:pvm}

{ \footnotesize
\begin{tabular}{cccrccccc} \label{tab:2}
&&&&&& \\
$M_{S}$ ($M_{\odot}$) & $t_{s}$ ($t_{\odot}$) & $ T_{s} $(K) &  & & & $\tau_{y} (\tau_{ye})$      \\
&&&& && \\
0.15&\ldots  & 3020  & $|$  & 0.09 & 0.03  & 0.01 & 6$\cdot 10^{-3}$ & \\
0.25& \ldots & 3311 & $|$ & 0.17  & 0.05 & 0.02 & 0.01 &    \\
0.4&\ldots & 3715 & $|$  & 0.39 & 0.11 & 0.05 &0.02  &   \\
0.6&\ldots & 4365 & $|$  & 0.73  & 0.22  & 0.09  & 0.05 &  \\
0.8& 2.5 & 5011  & $|$  & 1.5 & 0.45 & 0.19 & 0.1 &   \\
0.9& 1.5 & 5370  & $|$  & 2 & 0.6 &0.25  & 0.13 &  \\
1& 1 &  5800 & $|$ & 2.65 & \boxed{0.79} & 0.33 & 0.17    \\
1.1 & 0.64 & 6166 & $|$  & 3.64 & 1.08 & 0.5 & 0.23 &  \\
1.2& 0.45 & 6607 & $|$  & 4.94 & 1.46 & 0.62 & 0.32 &  \\
1.3& 0.32 & 6918 & $|$  &5.65  &1.67  & 0.7 & 0.36 &  \\
1.4& 0.25 & 7244 & $|$  & 7.65 &2.27  & 0.96 & 0.49  & \\
1.5&  0.2 & 7586 & $|$  & 8.85 & 2.62 &1.1  & 0.57 &  \\
2& 0.075 & 9550 & $|$  &18.18  &5.39  &2.27  &1.16  &  \\
3& 0.025 & 12590 & $|$  & 59 & 17.5 & 7.36 & 3.77 &  \\
4& 0.012 & 15135 & $|$  & 120.5 &35.7  & 15 &7.7  & \\
6& 5$\cdot 10^{-3}$ & 19953 & $|$  & 278 & 82.4 & 34.8  & 17.8 & \\
8&  3$\cdot 10^{-3}$ &22387 & $|$  & 480.6  & 142.4 & 60 & 30.76  &  \\
10& 2$\cdot 10^{-3}$ & 25120 & $|$  &722.6  &214  & 90 &46.2  &  \\
\\
& & & $$ & 200 & 300 & 400 & 500 & $T_{b} (\text{K})$ \\
&&&&&& \\
\end{tabular} } \\ 

In the particular case of our carbon life (H, C, O), as it comes on  Earth:
\begin{equation*}
\begin{array}{l}
T_{s} =  T_{\odot} = 5800 \, \text{K} \\
t_{s} =  t_{\odot} = 10^{10} \, \text{years} \\
\epsilon = 0.017 \\
R_{s} = R_{\odot} =   6.88 \cdot 10^{8} \, \text{m}  \\
M_{s} = M_{\odot} =
1.99 \cdot 10^{30} \, \text{kg} \\
a =   1.49 \cdot 10^{11} \, \text{m} \sim d \\
m_{p} = m_{e} = 6 \cdot 10^{24} \, \text{kg} \\
T_{b} = T_{av. obs.} \sim 288 \, \text{K} \\
\tau_{ye} = 3.15 \cdot 10^{7} \, \text{s} \\
\end{array}
\end{equation*}
From (\ref{eq:g2}) we get 
\begin{equation*} 
<T>   \sim 279 \, \text{K}
\end{equation*}

The difference between the calculated $<T>$ and $T_{av. obs.}$ observed \cite{ref:peo}
is explained by the combined effect of albedo and greenhouse due to the atmosphere. This explains why in the column (300 K) and in the row $M_{s} = M_{\odot}$, we find  a year a little shorter than the terrestrial year.

\section{Atmosphere, yes and no}
\setcounter{equation}{0}
\setcounter{figure}{0}
Consider a protagonist planet equipped with the right amount of atoms, namely those that enter into the structure of the living organisms. Next consider the molecules that are the building blocks and which undergo chemical reactions all the time.
In fact a living structure is different from the ensemble of the inert constituents. The inert, or lifeless, constituents stay in the physical-chemical state of minimum Gibbs potential pertinent to a given $T$ and $P$, while the organic structure does not.
Among the constituents we have the asymptotic molecules. For carbon life they are H$_{2}$O and CO$_{2}$.

The photosynthesis cycle requires a small area occupied by the photosynthetic centers and a large area occupied by the asymptotic components (indicated by $\mathcal{S}_{1}$ and $\mathcal{S}_{3}$ in \S 3). 
Therefore a process of release and return to the centers must be set into operation. This is called transport. Transport may come by diffusion and convection. Diffusion in solids is negligible; so we need asymptotic molecules in liquid and gas form.
For carbon life on Earth  CO$_{2}$ is always gaseous, H$_{2}$O sits in the three phases. Question: it is important, or even
 necessary, that at least one of the asymptotic molecules should explore all the phases? The answer is most likely yes. And the reason is the following. Both H$_{2}$O and  CO$_{2}$ are greenhouse gases. Greenhouse gases are necessary to control both the average temperature and its oscillations. This will be shown numerically
in \S 5. H$_{2}$O creates greenhouse by direct molecular interaction with the photons, and 
creates albedo by indirect interaction. Greenhouse belongs to quantum dynamics, albedo belongs to classical optics.
These interactions are linked by a triple feedback scheme as follows. 
\begin{center}
\includegraphics[width=6 cm]{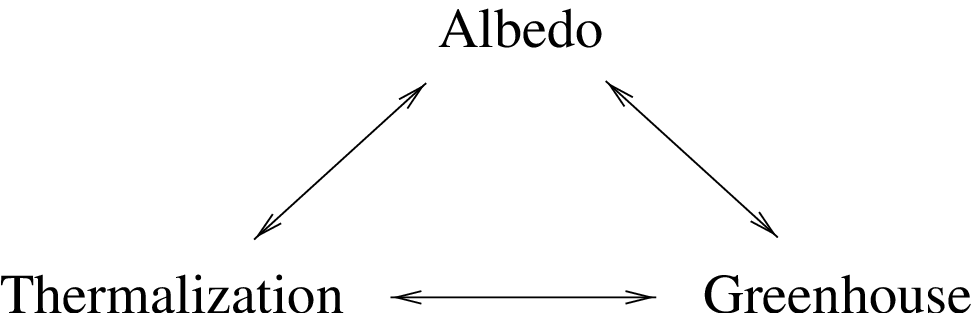} 
\end{center}
In fact H$_{2}$O evaporates from the liquid phase in bulk when the temperature $T (\vec{r}, t)$ and the pressure
$P (\vec{r}, t)$ command it. So H$_{2}$O moves to the atmosphere. Up there operates as a gas, and does the job of greenhouse.
But the temperature and pressure fields in the atmosphere change from place to place and from  time to time, so that the
H$_{2}$O sitting in the atmosphere may find itself in a local condition very close to condensation; it begins to develop a
droplet, evaporates, condenses again and so on. That is the H$_{2}$O in the appearance of cloud.
 A little cooler and the droplet becomes a drop and rains down; a little warmer and  H$_{2}$O  remains as a gas; 
in a right temperature band, and relative temperature fluctuations, we get a cloud; hence  albedo. 
Clouds have double effect: reduce the greenhouse because they steal H$_{2}$O molecule from the gas phase, moreover
 reduce by albedo the incoming radiation reaching the ground.
Less ground temperature means more super-saturation of H$_{2}$O in the air, and rain, and with it latent heat released. But this is only one step in the global water cycle.
Water in bulk travels and evaporates  elsewhere feeding again the greenhouse job. This complex circulation of mass and energy is self-stable.
The triple feedback is a powerful governor for both the inorganic dynamics and the life dynamics. In conclusion the transport of the asymptotic molecules \emph{must} be gaseous and \emph{should} also be liquid.

The feedback belongs to the realm of control dynamics and is, by definition, opportunistic. Therefore in this paper which is addressed to extrapolations and predictions regarding unknown combinations star-planet we cannot take into consideration opportunistic issues.

Bearing in mind these recommendations, we can examine, in the next section, the behaviour of the temperature
of the protagonist planet, which is assumed without atmosphere and consequently without pressure.
This theoretical planet should be used as a theoretical input, rather than a description of the  host  living planet.

\section{Temperature oscillations and total angular momentum}
\setcounter{equation}{0}
\setcounter{figure}{0}
We assume for the sake of simplicity that the system star-planet is isolated, and consequently, that the angular momentum
\begin{equation*}
\vec{J} = \vec{L} + \vec{\sigma}
\end{equation*}
where $\vec{L}$ is the orbital angular momentum and $\vec{\sigma}$ is the proper planet spin, is a constant vector.
The radiation flux impinging on the surface of the planet at a point P, depends on the kinematics star-planet; and the constraint $\vec{J} =$ const. alone is compatible with several possible configurations in coordinate space, each depending on the initial condition
in the star-planet phase space. Nobody gives the initial condition for any two body astronomical system.
We should rather think of the evolution of  each planetary system and consider that on the long time scales
the evolution can set 
itself in some sort of near stable dynamical configuration. These global configurations may be
 compatible with planetary 
elliptic orbits. This happens with the solar system, and it may be a general feature.

We consider our protagonist planet in this regime and we add another simplifying hypothesis, namely that both 
$\vec{L}$ and $\vec{\sigma}$ are conserved. The particular values of $\vec{L}$ and $\vec{\sigma}$, of course, come 
with the evolution of the global planetary system.
If we do not introduce this assumption, there is very little we can say analytically.

In conclusion, using $\vec{L}$ and $\vec{\sigma}$ fixed, we see in appendix B, that the flux $\phi^{in}$ has the form
\begin{equation} \label{eq:g5}
\phi^{in} 
= K (\psi, \epsilon) \, \mathfrak{f} (\theta, \varphi, \psi; \zeta, \psi_{0}, \omega_{d})
\end{equation}
where
$\psi$ is the time-angle on the ellipse, $0 \le \psi \le 2 \, \pi$ (fig. \ref{fig:gal3}); \\
 $\theta$ is the colatitude on the planet's surface 
$0 \le \theta \le \pi$; (fig. \ref{fig:spin}) \\
$\varphi$ is the longitude
$0 \le \varphi \le 2 \, \pi$; \\
 $\omega_{d} = \frac{2 \, \pi}{\tau_{d}}$  is the pulsation of the spin motion with given period $\tau_{d}$ (the
index $d$ stands for day).

A given meridian rotates about the fixed spin axis with fixed angular velocity $\omega_{d} $
\begin{equation}
\varphi = \omega_{d} \, t
\end{equation}
Notice that $t$ is related to $\psi$ by the time-angle equation $t = F (\psi)$, (\ref{eq:s23}),
 so that
\begin{equation} \label{eq:tt1}
\psi =  F^{-1} (t)
\end{equation}
The dependence  (\ref{eq:tt1}) of the orbital angle $\psi$ on $t$ is nonlinear, and implicit from (\ref{eq:s23}).
We cannot write $\psi = \omega_{y} \, t$ except in the $\epsilon = 0$ orbits. The angle $\psi$
is given mod $2 \, \pi$ and the function $t = F (\psi)$ is given mod $\tau_{y}$. Accordingly, we can define (appendix A)
\begin{equation*}
 \omega_{y} = \frac{2 \, \pi}{\tau_{y}}
\end{equation*}
with 
\begin{equation*} 
\tau_{y} =  2 \, \pi \, \sqrt{\frac{a^{3}}{\gamma \, (M_{S} + m_{p})}}
\end{equation*}
Using the explicit relationship between $\psi$ and $t$, $t = F (\psi)$ , we adopt  $\psi$ as the independent
variable in the calculation of $\phi^{in}$. \\
\paragraph*{Comments on $\omega_{d}$, $\omega_{y}$} 
The orbital motion is periodic with $\tau_{y}$;  this is the motion of the center of the planet in the 
center of mass system star-planet. On the contrary the motion of a point P (appendix B) is not periodic in general.
More precisely the motion of P is: \\
\quad \\
\begin{tabular}{ll}
 periodic with period $\tau_{y} = n \, \tau_{d}$,& if $\frac{\omega_{d}}{\omega_{y}} = n$ \\
periodic with period $n_{2} \, \tau_{y} = n_{1} \, \tau_{d}$, & if $\frac{\omega_{d}}{\omega_{y}} = \frac{n_{1}}{n_{2}}$ \\
non periodic, &  if $\frac{\omega_{d}}{\omega_{y}} = $ real number \\
\end{tabular} \\
\quad \\
The three cases are indicated in appendix A (fig. \ref{fig:gx6}).

Using as independent variable $\psi$ rather than $t$ the difficulties related to $\frac{\omega_{d}}{\omega_{y}}$ are carried
into the function $\mathfrak{f}$, or, more precisely, into the way in which the point P explores the domain 
 $\vec{d} \cdot \vec{n} > 0$ (appendix B, fig. \ref{fig:gx7}).

This is the general kinematics we are starting with. Given the input $\phi^{in}$, the general non equilibrium thermodynamics of the
planet follows. In general the fluid circulation on the surface layer of the planet has three main inputs:
\begin{description}
\item[-] thermal exchanges with the solid  surface which receives the radiation and thermalizes it,
\item[-] direct interaction with the radiation in the form of albedo,
\item[-] heating of the atmosphere due to greenhouse
\end{description}
Notice that the albedo in its turn is generated by  underlying phenomena in the thermalization domain, namely phase transitions
either permanent (ice) or going on and off (clouds). Consequently there is a complicated thermodynamical loop rather similar to a control feed back. 

The organic ecosystem participates to this feedback. So both
phase transitions and global life  are in some sort of ``dialogue''
with the star. This issue is outside the scope of this paper. Here we limit ourselves to the non equilibrium thermodynamics response to the stellar input in the first approximation, namely the Fourier equation (see Appendix C).
Notice that if we imagine a perturbative expansion of the  equations governing the 
 interaction between the fluid component and the rigid component,  the
  Fourier equation would be clearly the first order approximation. In this approximation, the temperature field follows the input linearly, except for a time delay \cite{ref:nc}.

After this long introduction, we return to the spin and orbit periodicity, namely $\omega_{d}$, $\omega_{y}$.
Notice that if $n_{1}$ is large and $n_{2}$ is small, the spin typical time $\tau_{d}$ is, in a certain sense, the dominating clock
which misses the period $\tau_{y}$ by an error which is a little percent  of $\tau_{y}$. So the important clock is $\tau_{d}$.
This happens with $\phi^{in}$, and with $T$ as well. 

We know that in the terrestrial realization of life the important clock is
$\tau_{d}$. Animals and human beings act along the time span of a day with accuracy of seconds or minutes.
On the contrary the phenomena recurrent with the year are met in general with inferior accuracy.
Clearly living organisms are driven by the daily attractor.

For a planet having a combination $\omega_{d}$ and $\omega_{y}$, with $n_{1}$  and $n_{2}$ both small, neither
$\tau_{d}$  nor $\tau_{y}$ can be good clocks. This poses difficult problems and, as terrestrial animals,
 we are unprepared to face them.

We limit ourselves to present a set of figures where we compare a planet that has the same $\zeta \sim 23^{\circ}$ as the Earth and the same value of the eccentricity, $\epsilon = 0.017$, but with
\begin{equation*} 
\omega_{d} = 6 \, \omega_{y}
\end{equation*}
instead of the canonical $\omega_{d} = 365 \, \omega_{y}$.

For this low value of $\epsilon$ there is quasi symmetry between the north and the south hemispheres.
 In other words north and south are simply a way of labelling the stars that a terrestrial 
observer can see. For large $\epsilon$ north and south are not at all symmetric, therefore their meaning is no longer
arbitrary; they are in fact thermodynamically different.

Furthermore low $\zeta$ means small polar cups. Remember that the definition of the polar cups is the following:

\begin{equation*}
0 < \theta < \zeta \quad \text{North} \quad
\begin{array}{ll}
\phi^{in} \ne 0; & \psi_{SE} < \psi < \psi_{AE} \\ 
\phi^{in} = 0; & \psi_{AE} < \psi < \psi_{SE} \\ 
\end{array}
\end{equation*}
\begin{equation*}
\pi - \zeta < \theta < \pi \quad \text{South} \quad
\begin{array}{ll}
\phi^{in} = 0; & \psi_{SE} < \psi < \psi_{AE} \\ 
\phi^{in} \ne 0; & \psi_{AE} < \psi < \psi_{SE} \\ 
\end{array}
\end{equation*}
SE: spring equinox, AE: autumn equinox. \\
The region $\zeta < \theta < \pi - \zeta$ contains the oscillation of the incoming flux with pulsation $\omega_{d}$.
This is the region where the daily clock operates.

Figures labelled with a), show the input $\phi^{in}$, figures with b) show the daily average input $\overline{\phi}^{in}$
\begin{equation}
\overline{\phi}^{in} = \frac{1}{2 \, \pi} \, \int_{0}^{2 \, \pi}  \, \phi^{in} \, \ud \varphi
\end{equation}
Figures labelled with c) show the temperature field for the rigid black planet, numerical solution
of the Fourier equation (appendix D). For $\epsilon = 0.017$, we see that the equinoxes
(indicated with E) and solstices (indicated with S) are almost equidistant.

\begin{figure}[ptbh]
\begin{center}
\includegraphics[width=16 cm]{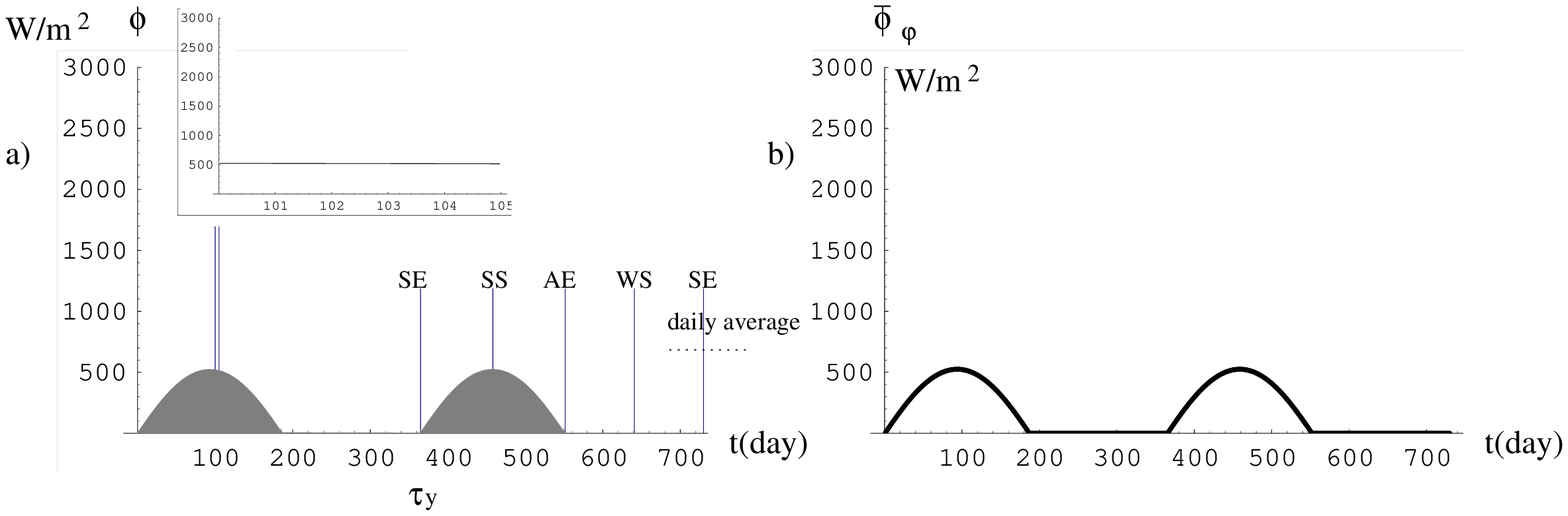} 
\includegraphics[width=9 cm]{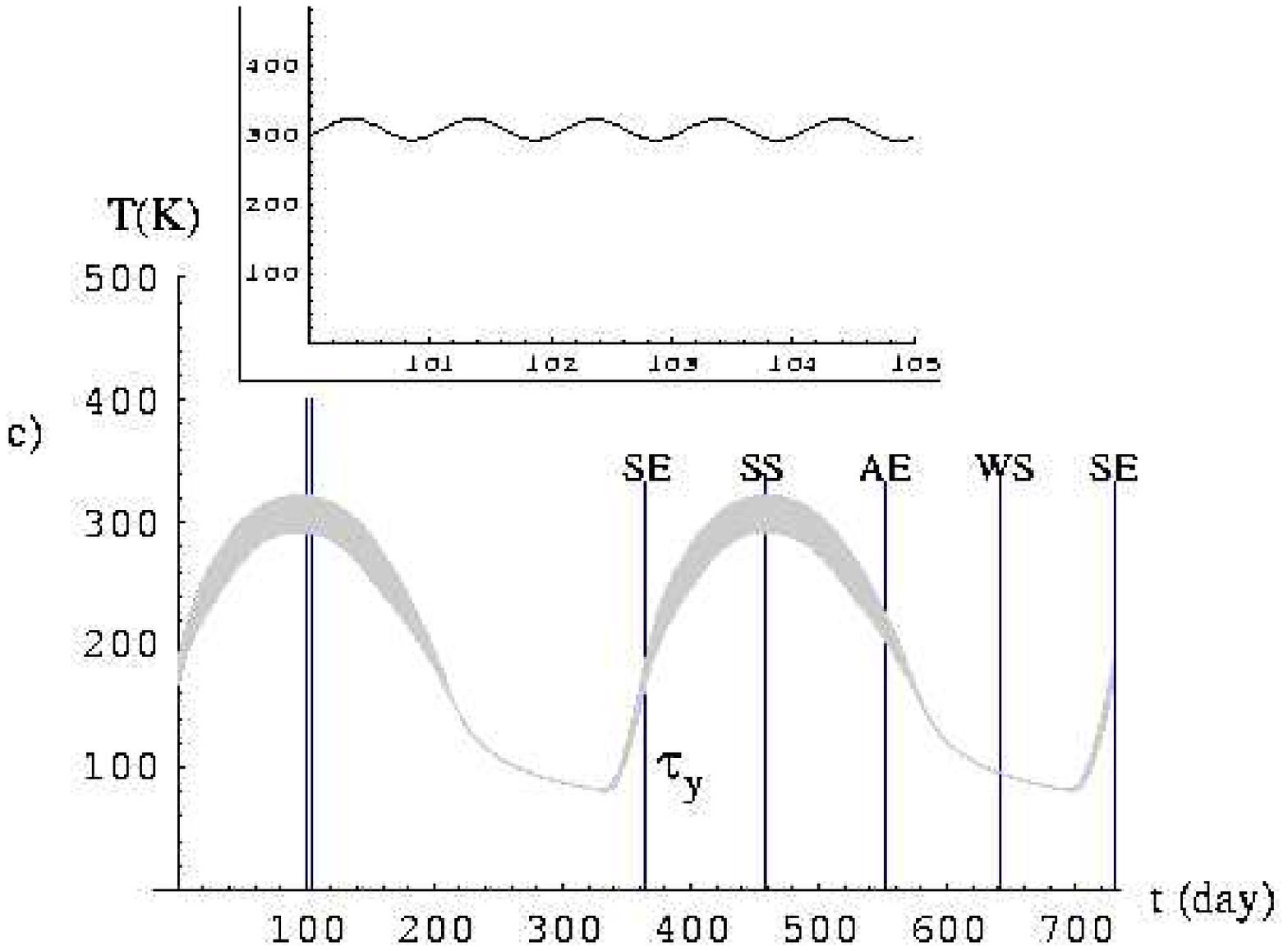}
\caption{  { \Large $\zeta = 23^{\circ}, \, \epsilon = 0.017, $ } 
\emph{ {\Large $  \omega_{d} = 365 \, \omega_{y}, \,
\theta = 0$. North pole.  
 }}}  \label{fig:gxx15}
\end{center}
\end{figure}

\begin{figure}[ptbh]
\begin{center}
\includegraphics[width=16 cm]{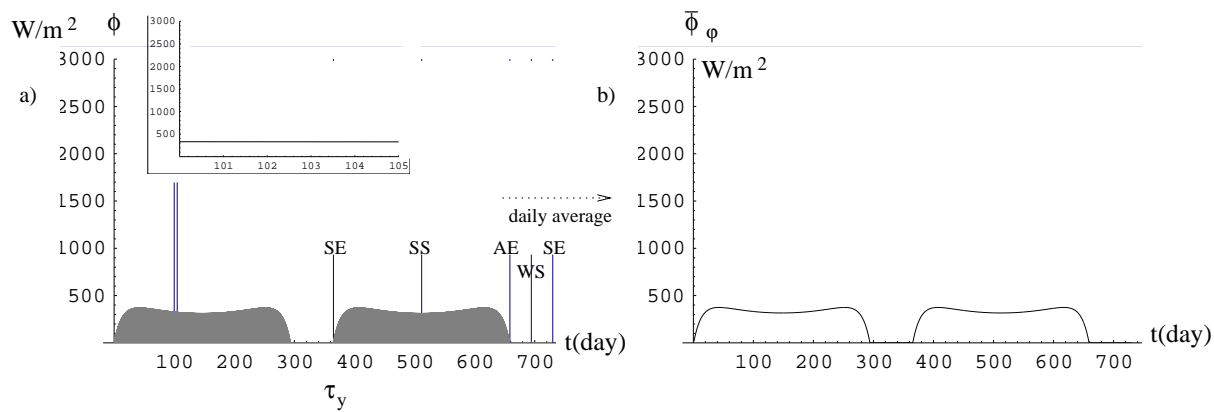} q
\includegraphics[width=9 cm]{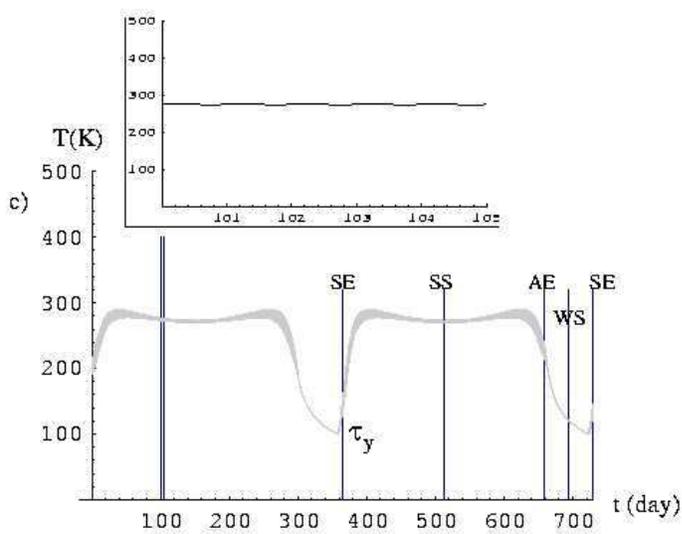} 
\caption{ \emph{ {\Large
$\zeta = 67^{\circ}, \, \epsilon = 0.5, \, \omega_{d} = 365 \, \omega_{y}$, $\theta = 0$. North pole.
 }}}  \label{fig:gx15}
\end{center}
\end{figure}

\begin{figure}[ptbh]
\begin{center}
\includegraphics[width=16 cm]{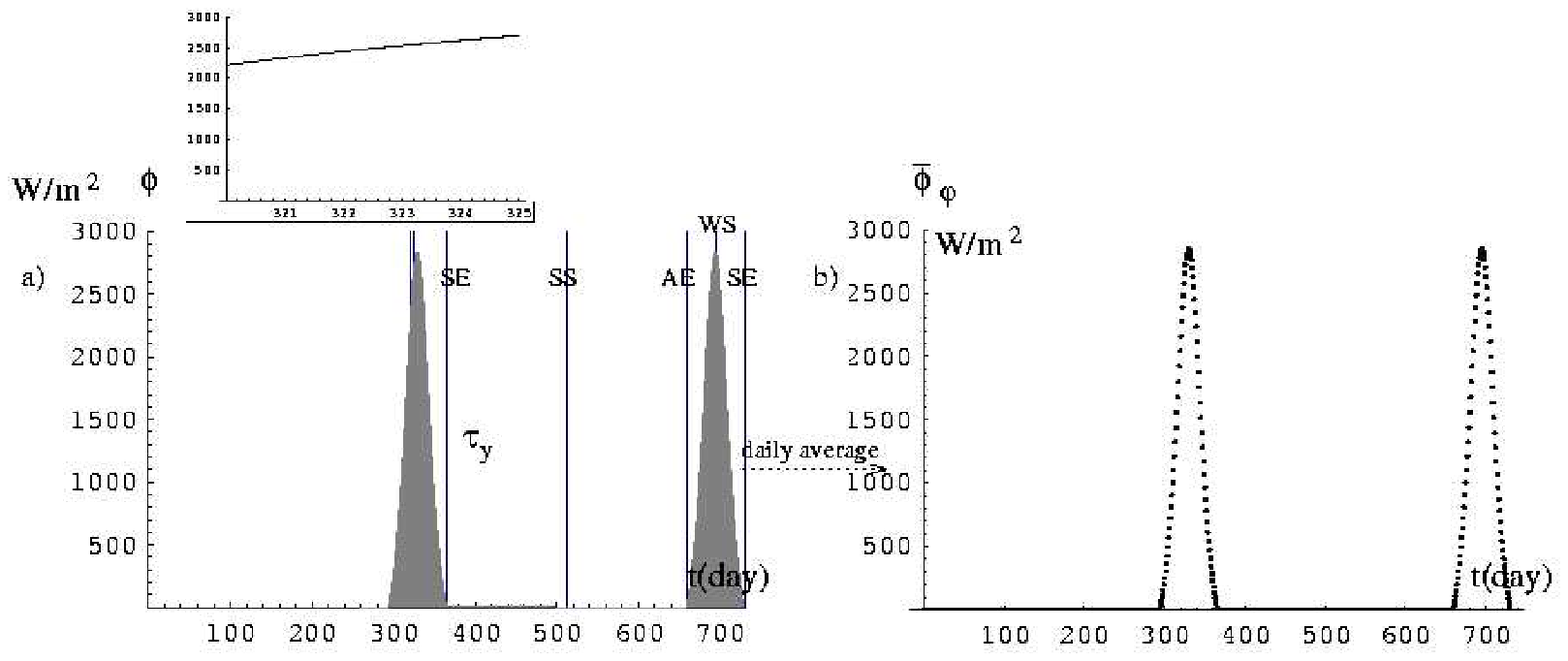} 
\includegraphics[width=9 cm]{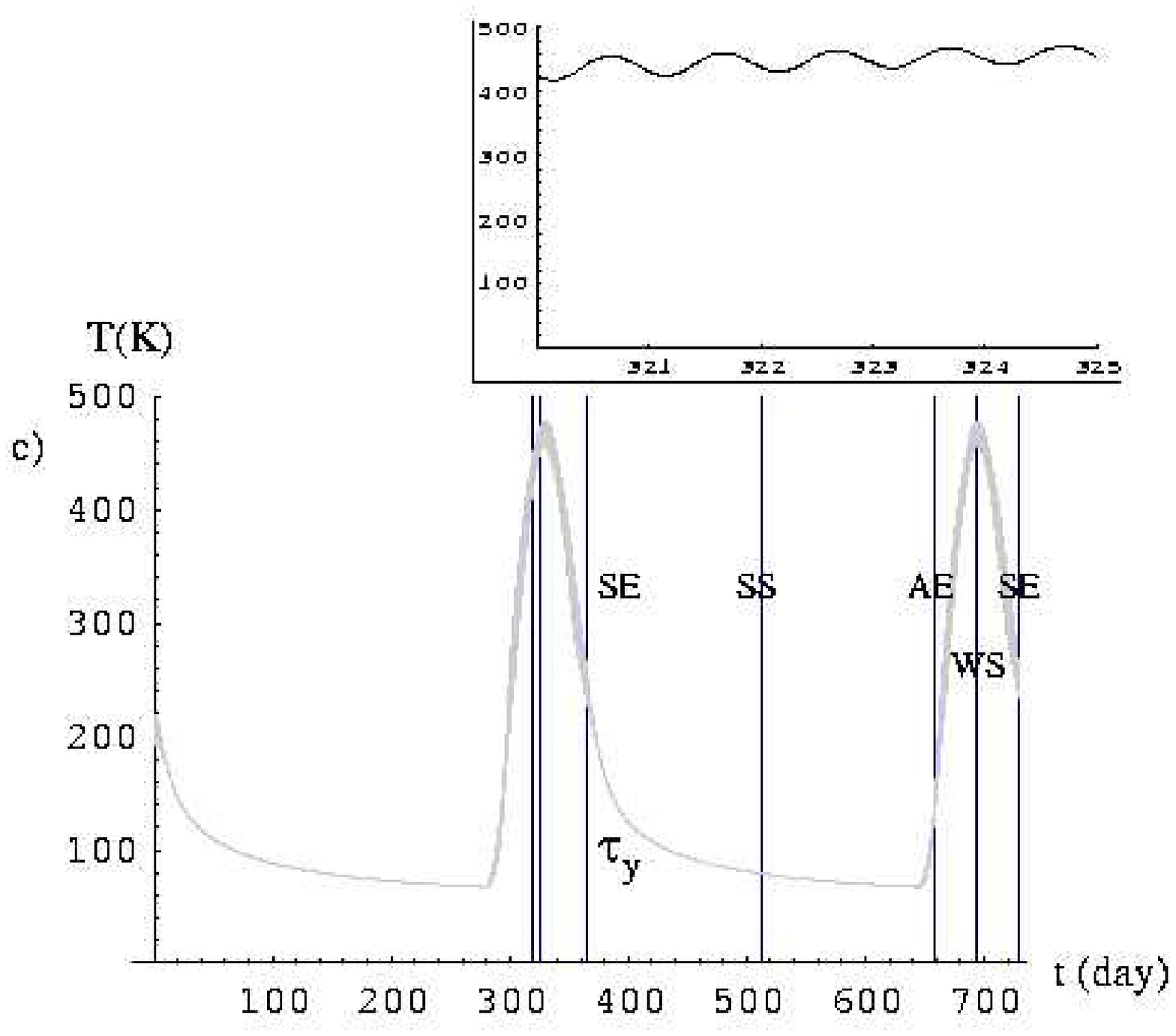} 
\caption{ \emph{ {\Large 
 $\zeta = 67^{\circ}, \, \epsilon = 0.5, \, \omega_{d} = 365 \, \omega_{y}$, $\theta = \pi$. South pole.
 }}}  \label{fig:gx150}
\end{center}
\end{figure}

In these figures the daily oscillations are so dense, that are not  distinguishable in the graphs.
We have indicated  with light grey when the input and temperature field contain daily oscillations.
Dark grey as in fig. \ref{fig:gxx15} a), \ref{fig:gx15} a), \ref{fig:gx150} a)  means no daily oscillations.

\begin{figure}
\begin{center}
\includegraphics{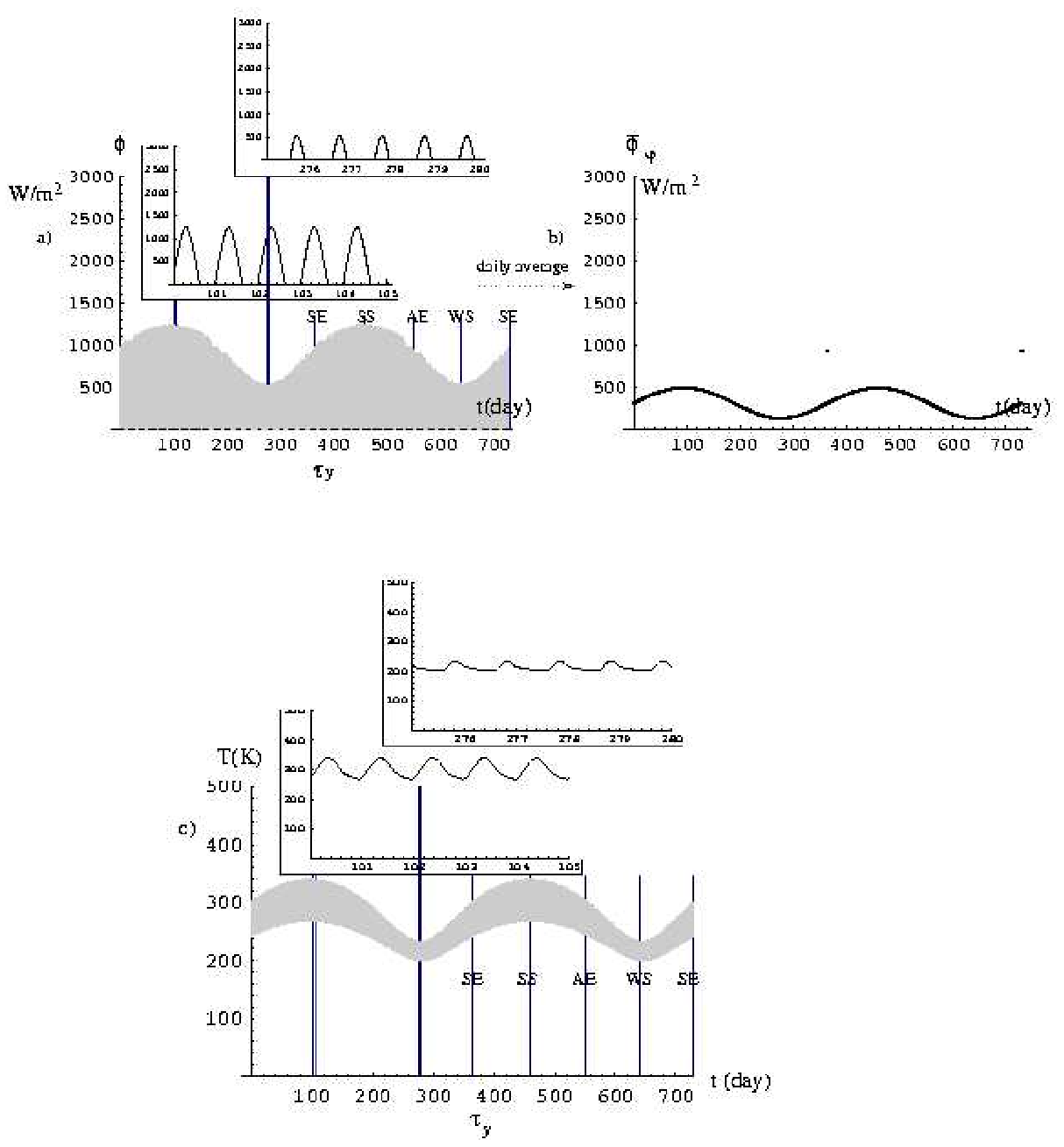} 
\vspace{-3cm} 
\caption{\emph{ { \Large $\zeta = 23^{\circ}, \, \epsilon = 0.017, \, \omega_{d} = 365 \, \omega_{y}, \,
\theta = \frac{\pi}{4}$. 
 }}}  \label{fig:gxx14}
\end{center}
\end{figure}

\begin{figure}
\begin{center}
\includegraphics{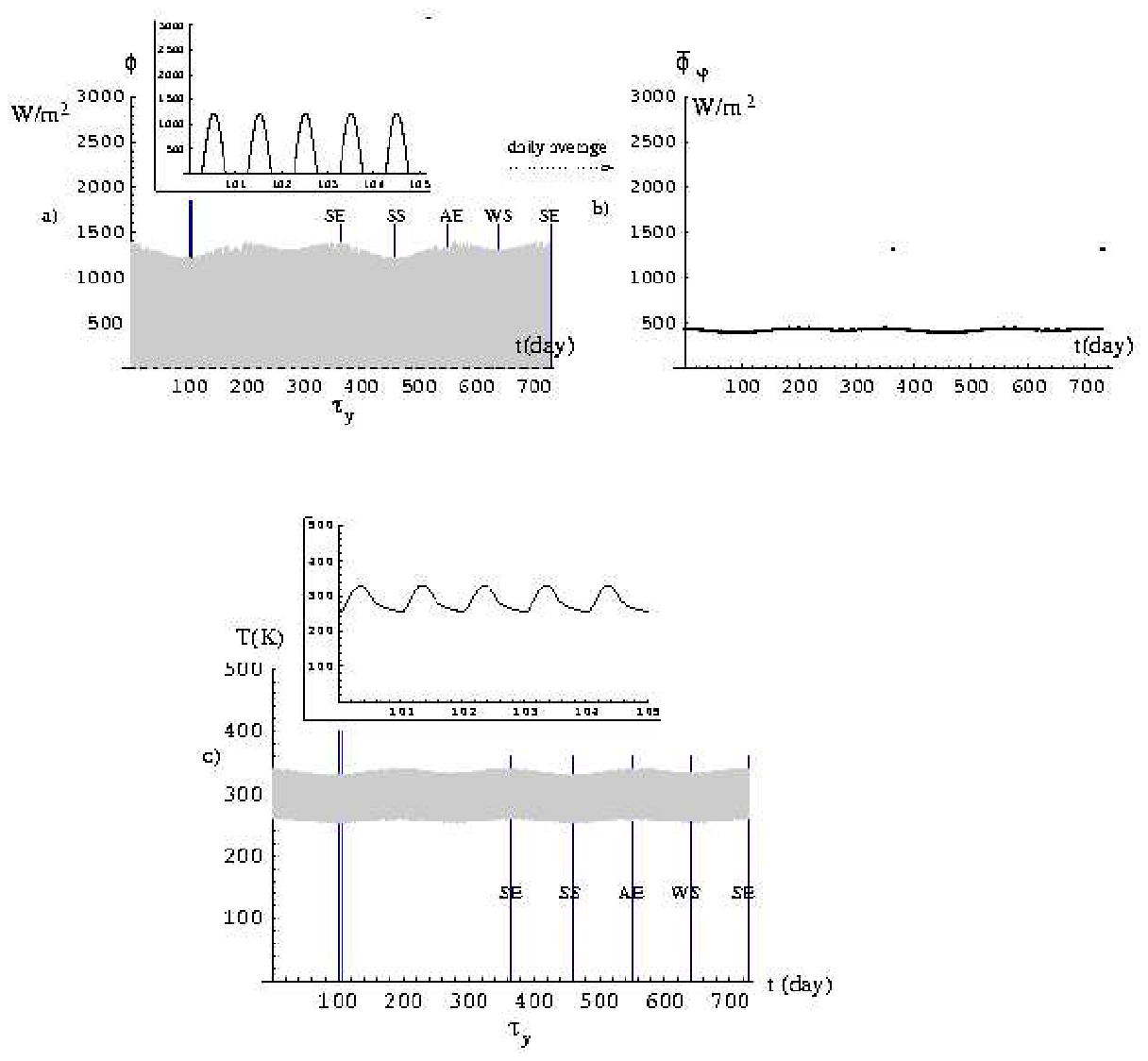} 
\vspace{-3cm} 
\caption{ \emph{ { \Large $\zeta = 23^{\circ}, \, \epsilon = 0.017, \, \omega_{d} = 365 \, \omega_{y}, \,
\theta = \frac{\pi}{2}$. Equator.
 }}}  \label{fig:gxx16}
\end{center}
\end{figure}

To familiarize with the graphs showing the calculated temperature field of the black planet, 
   we present in figures \ref{fig:gxx15},
\ref{fig:gxx14}, \ref{fig:gxx16} the case $\boxed{\zeta =
23^{\circ}, \, \epsilon = 0.017, \, \omega_{d} = 365 \, \omega_{y}}$, namely the Earth. These figures refer to $\theta = 0$, north pole,
$\theta = \frac{\pi}{4}$, and $\theta = \frac{\pi}{2}$, equator. The south hemisphere is not shown since for $\epsilon \sim 0$
it is the complement of the north hemisphere.

Fig. \ref{fig:gxx15} shows the north pole. 
Fig. a): in one year the flux is like one day (summer, the grey bump) and one night (winter, the flat $\phi^{in} = 0$ interval).
The length of the day is equal to the length of the night.  
The temperature is shown in fig. c). Evidently, the non zero temperature during the time in which the radiation is absent is due by Fourier conduction flowing from the  $\theta > \zeta$ zone. The small temperature oscillations during the time $\tau_{d}$
are also originated from the $\theta > \zeta$ zone. 
 Our graph helps understanding the role of the water phase transitions
on the thermal behaviour of the planet, comparing this graph with the observed temperature.

Fig. \ref{fig:gxx14} shows $\theta = \frac{\pi}{4}$.
This latitude is for the Earth the temperate zone. We see that both input and temperature 
have during the year two dips and two bumps. They are what by tradition we call four seasons.
The reader living at this latitude can look at the graphs and appreciate how much, and in which way, albedo, greenhouse,
wind and rain affect his own surrounding thermodynamics.

Fig. \ref{fig:gxx16} shows the equator.

\begin{figure}[ptbh]
\begin{center}
\includegraphics[width=14 cm]{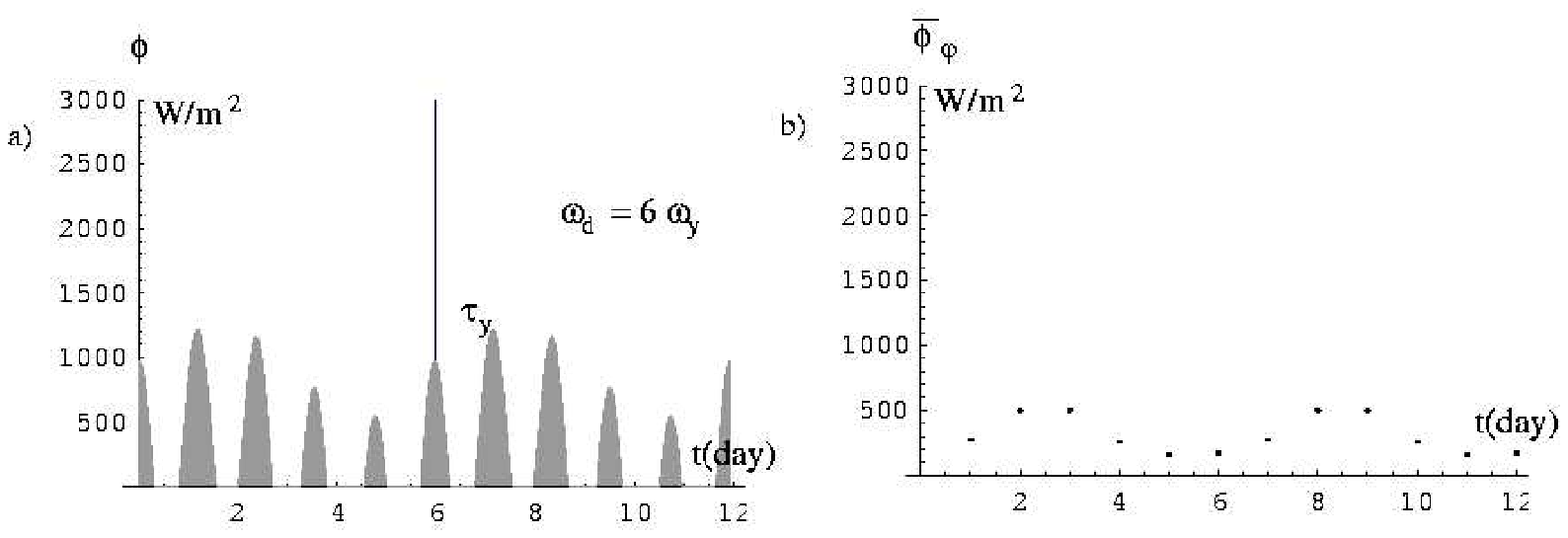} 
\includegraphics[width=9 cm]{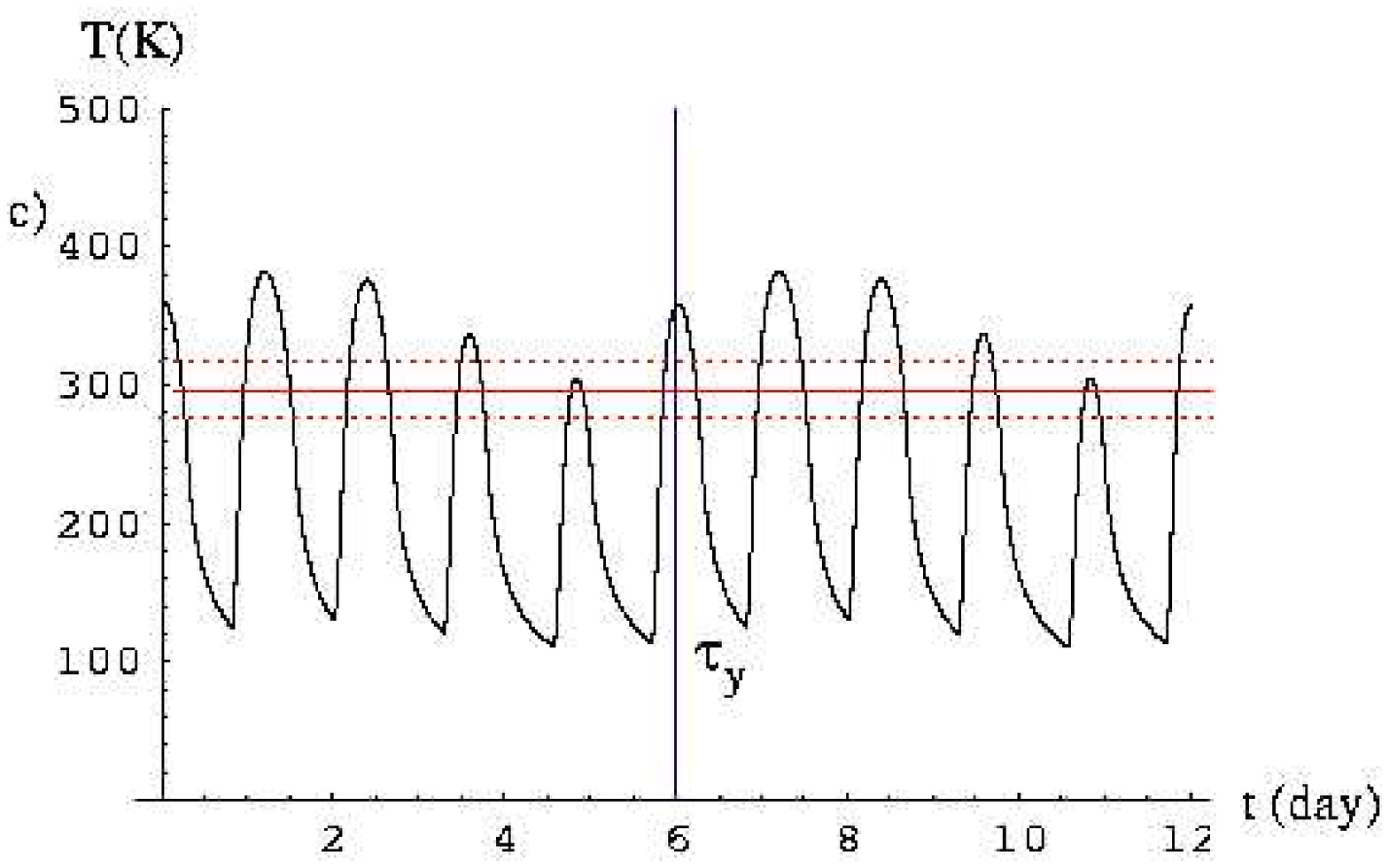} 
\caption{ \emph{ {\Large $ \zeta = 23^{\circ}, \, \epsilon = 0.017, \,
\omega_{d} = 6 \, \omega_{y}$, $\theta = \frac{\pi}{4}$.  
 }}} \label{fig:gx18}
\end{center}
\end{figure}

We move then to the comparison with the case $\zeta =
23^{\circ}, \, \epsilon = 0.017$ but $\boxed{ \omega_{d} = 6 \, \omega_{y}}$ instead of $ \omega_{d} = 365 \, \omega_{y}$.
Fig. \ref{fig:gx18} shows $\theta = \frac{\pi}{4}$. 
We see first of all that the concept of season is hardly visible, by considering fig. a). The temperature has a huge amplitude of oscillation and a characteristic difference between the cooling branch and the heating branch connected to a stronger time delay between temperature and radiation input in the descending branch. This property has been discussed in
 \cite{ref:nc}.  See also appendix C.
Greenhouse and air circulation in general act in the direction of depressing the oscillations that are proper to the Fourier equation.

In general, we can say that both a high daily frequency and the fluid circulation act together as  a mixer that homogenizes the temperature
field on the surface. 

In conclusion we need acrobatic hypotheses on the fluid circulation to modify the above oscillations which
 are incompatible with the request of $T_{b}$ lying in a reasonable band. See fig. 
\ref{fig:gx18} c).

\paragraph*{Comments on $\zeta, \, \epsilon$}
We keep for the semi-major axis $a$ the  same value that we have for Earth and we keep also $\omega_{d} = 365 \, \omega_{y}$
in order to have something to compare with, when we study a different value of $\zeta$ and $\epsilon$.

Notice that same $a$ means same $\tau_{y}$, and same ratio $\frac{\omega_{d}}{\omega_{y}}$ means same mixing power of homogenization
on the temperature filed, as discussed above. In the following figures we illustrate the case 
$\boxed{\zeta = 67^{\circ}, \, \epsilon = 0.5, \, \omega_{d} = 365 \, \omega_{y}}$. First of all with $\epsilon = 0.5$ the symmetry north-south is completely lost.
Solstices and equinoxes are unequally separated from each other. There is a long interval between spring equinox and autumn equinox and a short interval between autumn equinox and the next spring equinox.

Fig. \ref{fig:gx15} represents north pole: a long warm summer and a short  extremely cold winter. At the south pole we get instead (fig. \ref{fig:gx150}) a very long cold summer  and a short hot winter. 

\emph{Notice that on the Earth the populations living in the South hemisphere are used to call ``summer'' the period between the autumn equinox and the next spring equinox, namely the interval of time that the populations living in the North hemisphere call ``winter''.
Here we do not adopt this inversion of the names.}

Moreover for $\zeta = 67^{\circ}$ the polar cups are large, the temperate zone is very narrow.

\begin{figure}[ptbh]
\begin{center}
\includegraphics[width=16 cm]{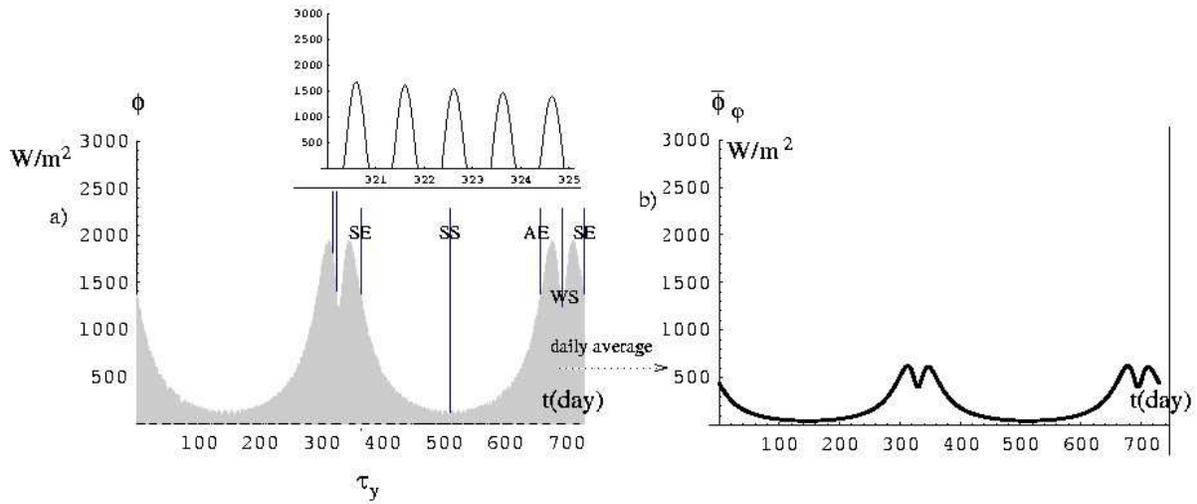} 
\includegraphics[width=9 cm]{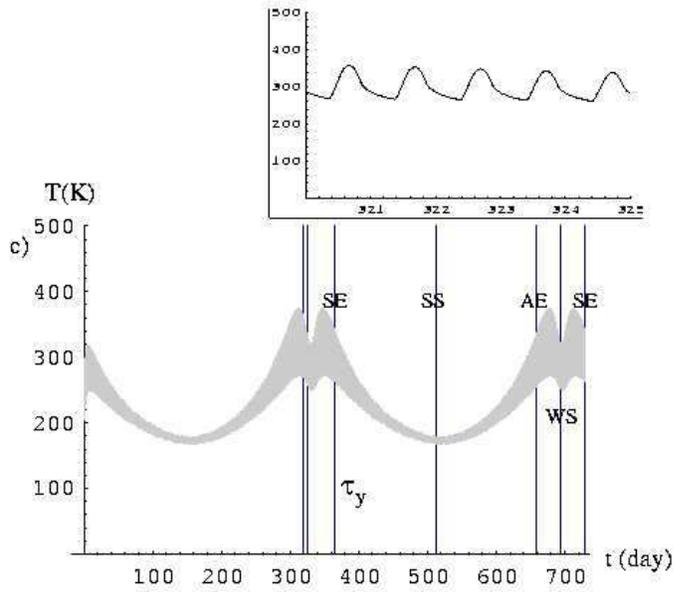} 
\caption{ \emph{ {\Large  
$\zeta = 67^{\circ}, \, \epsilon = 0.5, \, \omega_{d} = 365 \, \omega_{y}$, $\theta = \frac{\pi}{2}$. Equator.
 }}}  \label{fig:gx16}
\end{center}
\end{figure}

Let us consider the equator, fig. \ref{fig:gx16}. The seasons appear in a wild way. We have 4 seasons, namely two dips and two peaks
with a long cold summer and a short interval in which are packed the two equinoxes and the winter solstice.

\begin{figure}[ptbh]
\begin{center}
\includegraphics[width=16 cm]{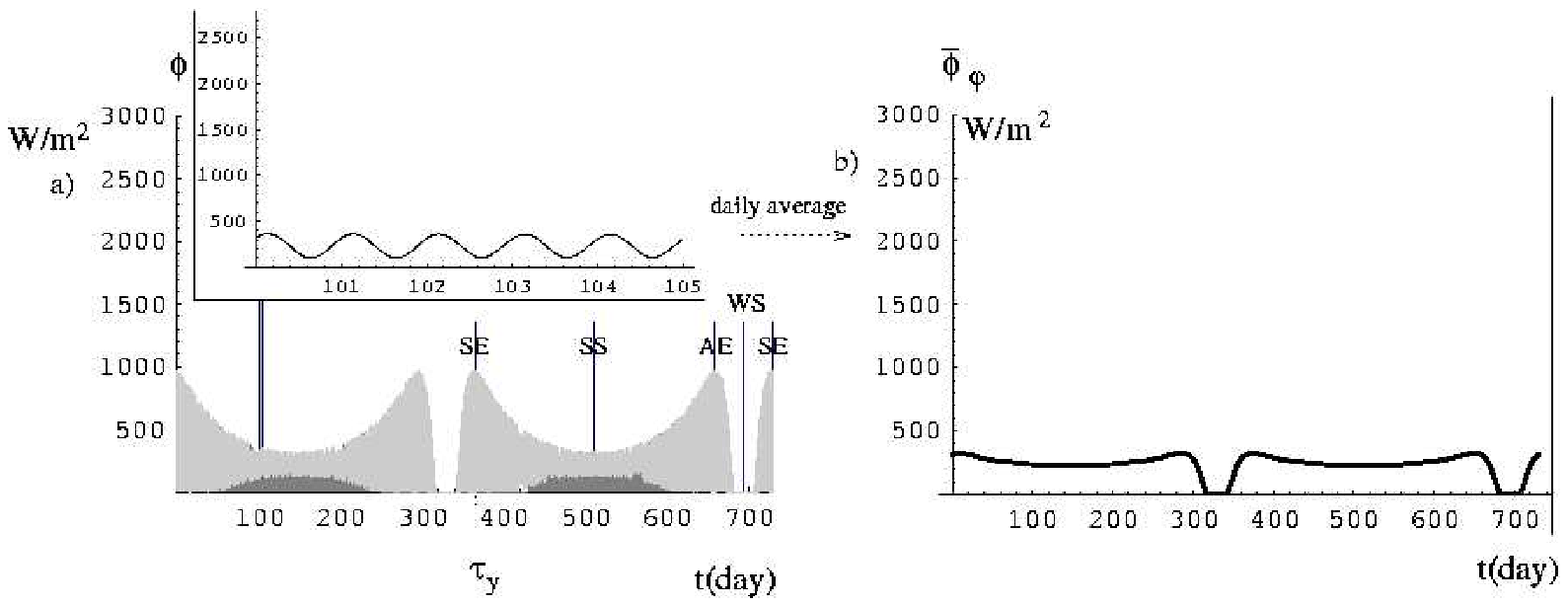} 
\includegraphics[width=9 cm]{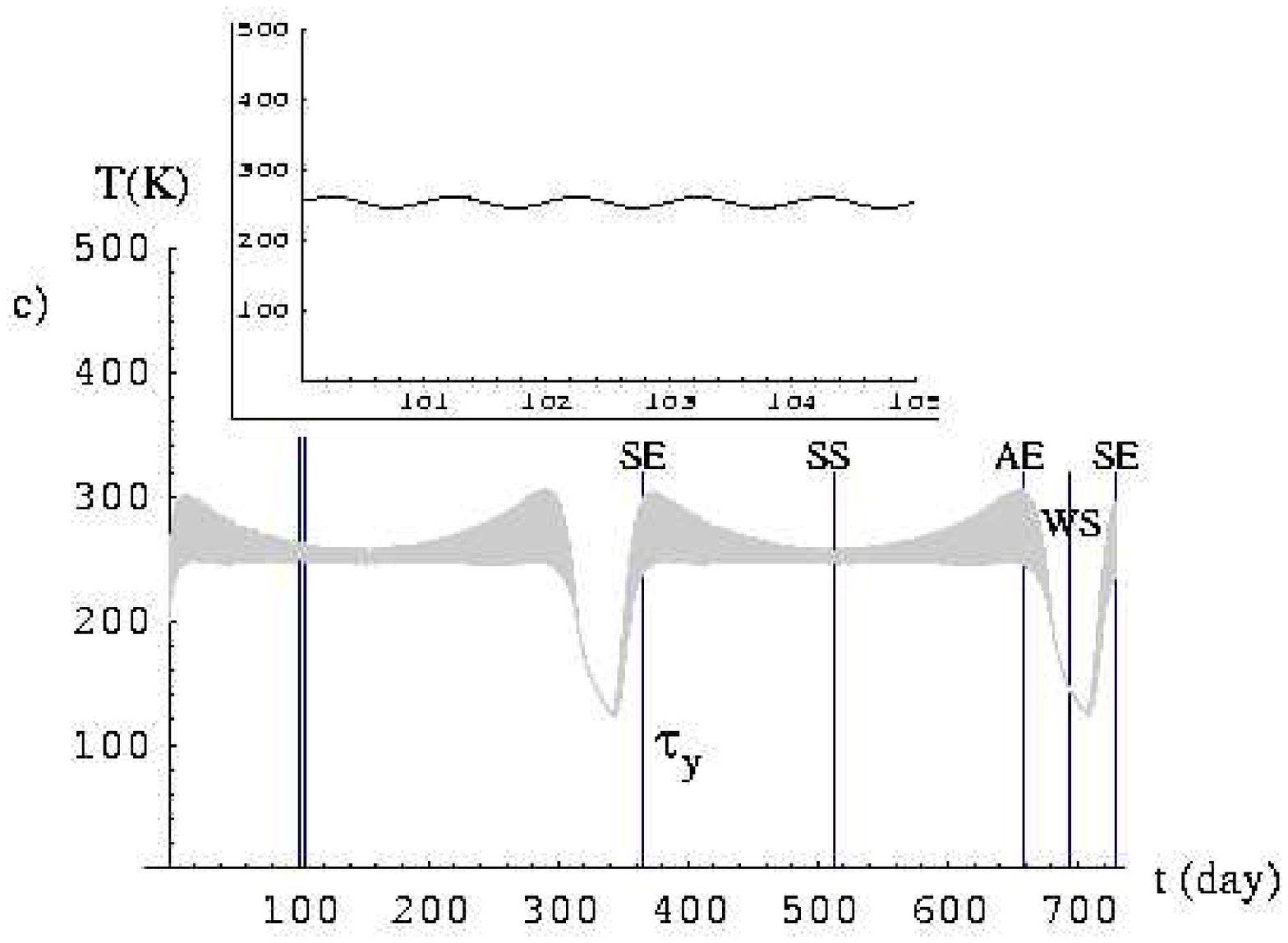} 
\caption{ \emph{ {\Large  
 $\zeta = 67^{\circ}, \, \epsilon = 0.5, \, \omega_{d} = 365 \, \omega_{y}$, $\theta = \frac{\pi}{4}$.
 } {\footnotesize Note: the meaning of the darker area in a) is explained in the magnified drawing on the upper part of the figure:
the incoming radiation oscillates with pulsation $\omega_{d}$, but there is not radiationless night}}}  \label{fig:gx14}
\end{center}
\end{figure}

\begin{figure}[ptbh]
\begin{center}
\includegraphics[width=16 cm]{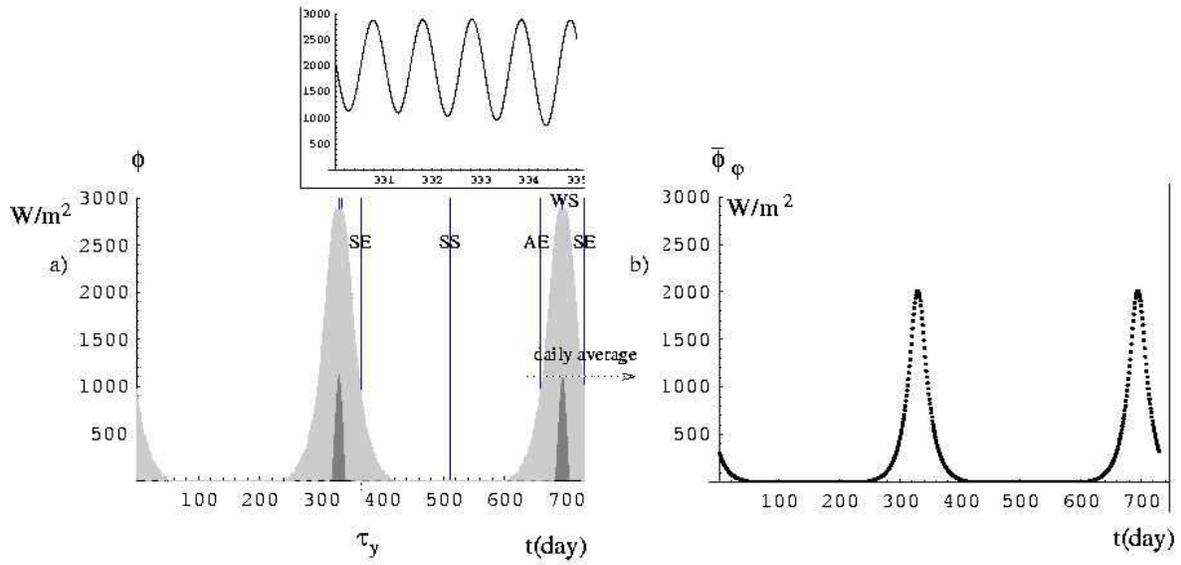} 
\includegraphics[width=9 cm]{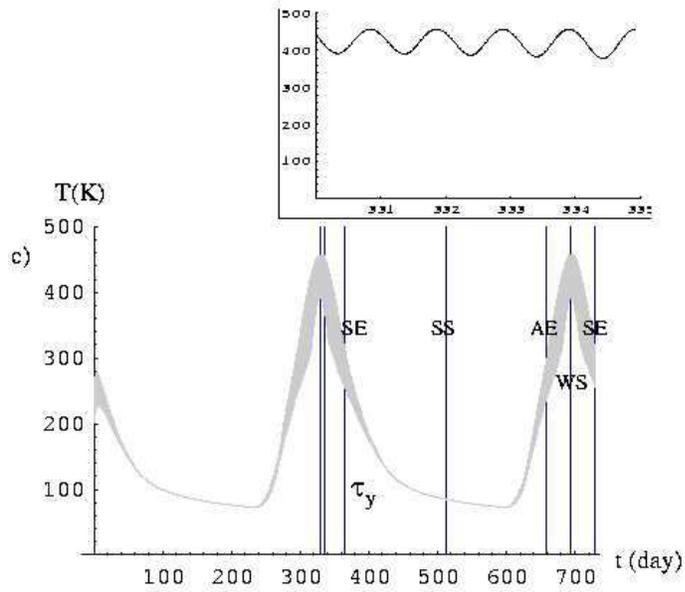} 
\caption{ \emph{ {\Large  
 $\zeta = 67^{\circ}, \, \epsilon = 0.5, \, \omega_{d} = 365 \, \omega_{y}$, $\theta = \frac{3 \, \pi}{4}$.
 }}}  \label{fig:gx140}
\end{center}
\end{figure}

The middle latitude $\theta = \frac{\pi}{4}$ and $\theta = \frac{3 \, \pi}{4}$ are shown in fig. \ref{fig:gx14} and \ref{fig:gx140}.
These latitudes that are temperate zones for the Earth; in this case instead are very close to the behaviour
 of the  poles. The temperate zone in fact is a narrow band around the equator.
 
In conclusion the orbital parameters $\zeta = 67^{\circ}, \, \epsilon = 0.5$ are highly unfit for life, unless acrobatic hypothesis on the atmosphere are injected into the general picture.
\newpage 
\newpage 

\section{Concluding remarks}
\setcounter{equation}{0}
\setcounter{figure}{0}
We have illustrated in \S 5 the cases $\omega_{d} = 365 \, \omega_{y}$ and $\omega_{d} = 6 \, \omega_{y}$.
We did so for ease of calculation and because the year's and day's cycles indeed dominate the surface thermodynamics of the planet
in the global stationary regime. Choosing two examples satisfying $\omega_{d} = n \, \omega_{y}$ is therefore significant.

On the other hand we remark that the terrestrial ratio $\frac{\omega_{d}}{\omega_{y}}$ is not exactly 365, and in general the rule 
$\frac{\omega_{d}}{\omega_{y}} = \frac{n_{1}}{n_{2}}$ cannot be expected to be the law of nature. In fact complicated astronomical
events are responsible for such ratio.

When  $\frac{\omega_{d}}{\omega_{y}} =$ real number, the temperature field is more complicated and we have shown in appendix C
that the average field $<T>_{\theta, \varphi} (t)$ has annual periodicity, in the stationary regime, but the field 
$T (R, \theta, \varphi, t)$ is non periodic. This means that the Fourier expansion of $T (R, \theta, \varphi, t)$ spans the infinite interval 
\begin{equation*}
0 < t < \infty
\end{equation*}
and therefore $T$ contains a continuous spectrum in the frequency $\nu$. Instead the average $<T>_{\theta, \varphi} (t)$
spans the finite interval
\begin{equation*}
t' < t < t' + \tau_{y} 
\end{equation*}
 and therefore $<T>_{\theta, \varphi} (t)$ contains a discrete spectrum:
\begin{equation*}
 \nu = \frac{M}{\tau_{y}} 
\end{equation*}
This strange fact comes as a consequence of the simplicity of the
elliptic motion, which determines for the radiation input the factorized form
\begin{equation}  \label{eq:x1}
\phi^{in} = K (t) \, \mathfrak{f} (\theta, \varphi, t)
\end{equation}
Evidently, a planetary motion sensible to many body interactions, produces a
 radiation input which  is much more complicated than  (\ref{eq:x1})  \cite{ref:mac}
and the temperature field $T (R, \theta, \varphi, t)$  is extremely difficult to analyze.

The above comments come from the fact that the equation of motion (see appendix C)
\begin{equation} \label{eq:z1}
c_{v} \, \rho_{v} \, \frac{\de T}{\de t} = 
  \kappa \,  \Delta_{r, \theta, \varphi} \, T + \phi^{in} (\theta, \varphi, t) \, \delta (r - R)
\end{equation}
is linear dissipative and is attracted by the driving term (\ref{eq:x1}).

We can make at this point a step forward and move to the consideration of the fluid motions on the surface of the planet, see \S 4.
The partial differential equation of the fluid motions, the Navier-Stokes equation containing the heat transport,
is non linear, and is coupled to the linear equation (\ref{eq:z1}) which acts as the free field, or unperturbed term,
in the self-consistent description of the interaction fluid-rigid surface.
The fluid motions  are driven by the field $T$ solution of (\ref{eq:z1}), and in turn operate on it according to some 
iterative procedure, not discussed in this paper.

Now, the fluid velocity field has the property of being chaotic.
If we consider the global solution, we are induced to think that it will manifest a regular part and a chaotic part in some sort of proportion. 
This concept of ``proportion'' is the extrapolation of the property discussed in this paper: periodic the surface average, non periodic locally.
It is exactly this proportion which is the most important constraint that the planetary orbit induces in the relationship
Star-Life.

The living organism must cope both with the regular, periodic part of the ambient, and with the remaining chaotic unpredictable part of the ambient. Therefore its control strategy must be quite complex, perhaps we may say intelligent, and able to deal
with predictable ambient inputs plus unpredictable ambient inputs.
This analysis could be rejected in favor of the principle that unpredictability is dictated by geography.
In this view the discussion of extraterrestrial life seems to be reduced to casualness, so that we return to the geocentric viewpoint. This is not the viewpoint adopted in this paper. Above the uncertainties of geography stand the general constraints discussed
in \S 5, where we show that there are acceptable or non acceptable combinations of star temperatures and orbit semi-axis, plus acceptable or non acceptable combination of eccentricity and inclination. The definition of constraint we have adopted rests solely on general thermodynamics, and this is the only way we have to talk of the compatibility with life on planets which are not directly observable.

\section*{Appendix A. The motion on the ellipse}
\renewcommand{\theequation}{A.\arabic{equation}}
\renewcommand{\thefigure}{A.\arabic{figure}}
\setcounter{figure}{0}
\setcounter{equation}{0}
Let us consider the ellipse of 
 fig. \ref{fig:gal3}. This drawing corresponds to eccentricity $\epsilon = 0.8$. In this drawing it is also indicated the projection of the spin $\vec{\sigma}$ in the orbital plane, $\vec{\sigma}_{a}$, $\psi_{0}$ is the orbital precession.
At point $F$ (the first focus) there is the star. At point $P$ there is the planet. 
\begin{figure}[ptbh]
\begin{center}
\includegraphics[width=8 cm]{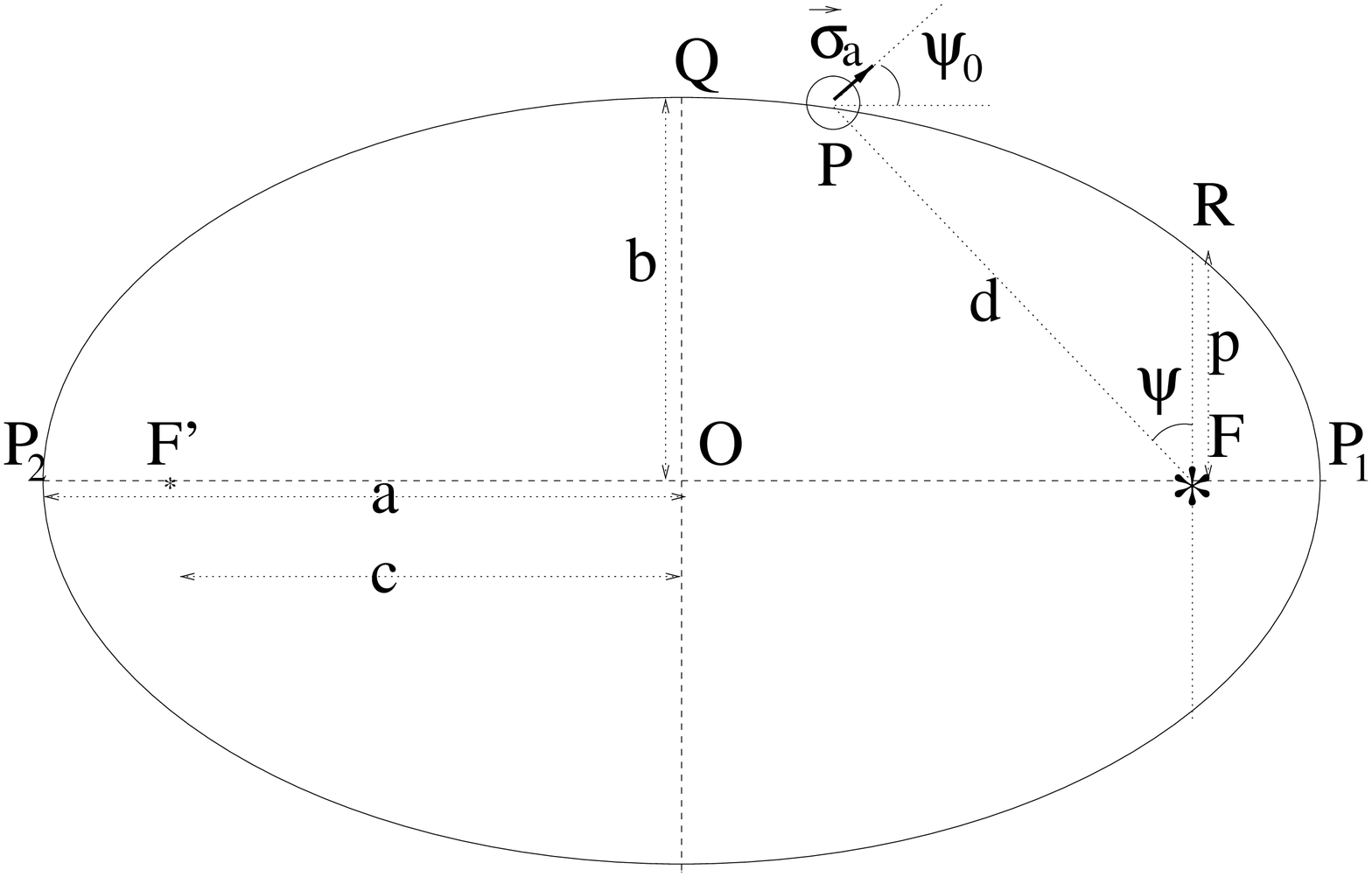}
\caption{ \emph{ {\footnotesize }}} \label{fig:gal3}
\end{center}
\end{figure} \\
$d, \, \psi$  are polar coordinates centered in $F$. 
\\
$d (\psi) = PF$ is the distance star-planet.
$P_{1}F = $ perihelion; \\
$P_{2}F = $ aphelion; \\
$P_{1}O = P_{2}O  = a = $ major semi-axis; \\
$FO = F'O = c = $ coordinate of the focus in the Cartesian system centered in $O$; \\
$\epsilon = \displaystyle{\frac{c}{a}} = $ eccentricity; \\
$OQ  = b = a \, \sqrt{1 - \epsilon^{2}} = $ minor semi-axis; \\
$RF = p = a \, (1 - \epsilon^{2}) $; 

\begin{figure}[ptbh]
\begin{center}
\includegraphics[width=12 cm]{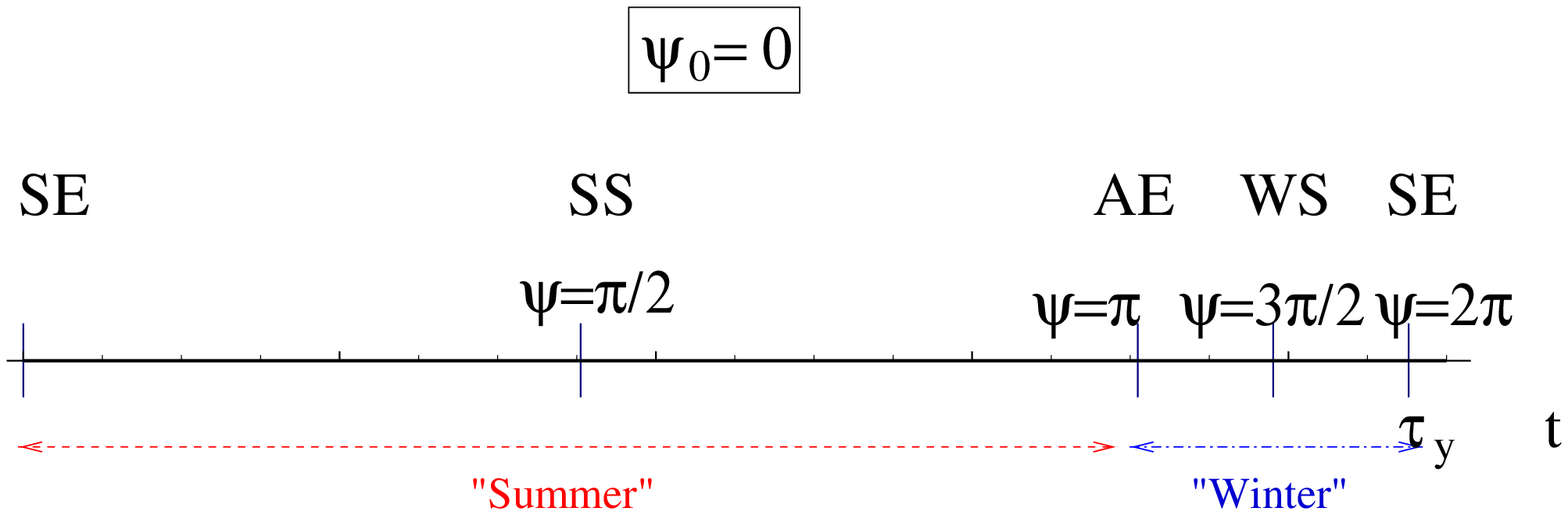} 
\includegraphics[width=12 cm]{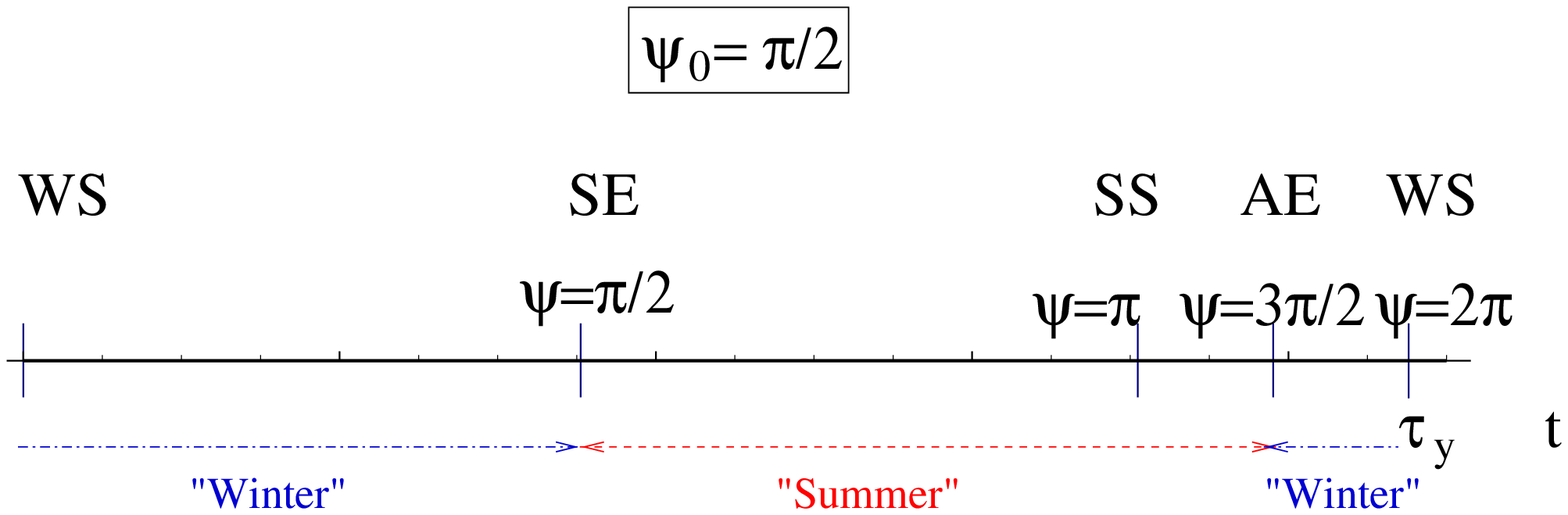} 
\caption{ \emph{ {\footnotesize The relationship time-angle $\epsilon = 0.5$. 
 }}}  \label{fig:gal4}
\end{center}
\end{figure}

The ellipse equation in the above polar coordinates is:
\begin{equation} \label{eq:s20}
d (\psi) = \frac{p}{1 - \epsilon \, \sin \psi} = \frac{a \, (1 - \epsilon^{2})}{1 - \epsilon \, \sin \psi}
\end{equation}
From (\ref{eq:s20}) we see explicitly the dependence on the parameters
 $a$  and $\epsilon$. In gravitational dynamics the physical parameters are
 $M_{s}$, star mass, $m_{p}$, planet mass, $\gamma$, Newton constant.
The equation of motion is:
\begin{equation} \label{eq:s21}
\mu \,  \, \ddot{\vec{\rho}} =
\frac{m_{p} \, M_{s}}{( m_{p} + M_{s} )}  \, \ddot{\vec{\rho}} =  \gamma \, 
\frac{m_{p} \, M_{s}}{\rho^{3}} \, \vec{\rho} 
\end{equation}
The solution of the vector equation of motion (\ref{eq:s21}) is a two dimensional trajectory lying in the plane orthogonal to $\vec{L} = \mu \, \vec{d} \wedge  \vec{v} $ (the ecliptic plane). The trajectory is determined once we fix 
$E = E_{p} + V(d)$, total energy, and $L = \mid \vec{L} \mid$.
The two parameters  $a$  and $\epsilon$, and the two constants of the motion $E$ and $L$, are related by the following equation:
\begin{equation} \label{eq:s22}
\begin{array}{ll} 
E & = \displaystyle{- \gamma \, \frac{m_{p} \, M_{s}}{2 \, a}} \\
& \\
L & = \displaystyle{ \sqrt{ \gamma \, a \, \mu \, ( 1 - \epsilon^{2} ) \, 
m_{p} \, M_{s}}} \\
\end{array}
\end{equation}
Since the motion obeys a time reversal invariant equation, the origin of the time is arbitrary. We set $\psi_{0} = 0$ and adopt this convention relating $\psi$ and $t$
\begin{equation} 
\begin{array}{llll} 
\psi = 0 & \text{spring equinox}   &  t = 0 & SE \\
& \\
\psi = \frac{\pi}{2} & \text{summer solstice}   &  t = t_{ss} (\epsilon) & SS \\
& \\
\psi = \pi & \text{autumn equinox}   &  t = t_{ae} (\epsilon) & AE \\
& \\
\psi = 3 \, \frac{\pi}{2}  & \text{winter solstice}   &  t = t_{ws} (\epsilon) & WS \\
& \\
\psi = 2 \, \pi & \text{spring equinox}   &  t = \tau_{y} & SE \\
\end{array}
\end{equation}
These names are related to the fact that for $\psi = 0$ the projection of the spin $\vec{\sigma}$ on the ecliptic,
$\vec{\sigma}_{a}$, is perpendicular to $\vec{d}$: 
\begin{equation*} 
(\vec{\sigma}_{a} , \vec{d}) = \frac{\pi}{2}
\end{equation*}
If we take into account the precession $\psi_{0}$, we get
\begin{equation*} 
(\vec{\sigma}_{a} , \vec{d}) = \frac{\pi}{2} + \psi_{0}
\end{equation*}
Keeping always the convention $\psi = 0, \, t = 0$, we have in the case of precession $\psi_{0}$
\begin{equation} 
\begin{array}{llll} 
\psi = \psi_{0} & \text{spring equinox}   &  t = t_{0} (\psi_{0}) & SE \\
& \\
\psi = \frac{\pi}{2} + \psi_{0} & \text{summer solstice}   &  t = t_{ss} (\epsilon) & SS \\
& \\
\psi = \pi + \psi_{0} & \text{autumn equinox}   &  t = t_{ae} (\epsilon) & AE \\
& \\
\psi = 3 \, \frac{\pi}{2} + \psi_{0}  & \text{winter solstice}   &  t = t_{ws} (\epsilon) & WS \\
& \\
\psi = 2 \, \pi + \psi_{0} & \text{spring equinox}   &  t  = t_{0} + \tau_{y} & SE \\
\end{array}
\end{equation}
(see also fig. \ref{fig:gal4}).
Notice that $\psi$ is the natural parameter that describes the elliptic motion. On the other hand the time $t$ is the natural
parameter for the description of the physical and biological dynamics. \\
The relationship between $\psi$ and $t$ is given by
\begin{equation*} 
t   = \frac{\mu}{L} \,
\int \, \rho^{2} (\psi) \, \ud \psi = F (\psi)
\end{equation*}
The explicit calculation gives
\begin{equation} \label{eq:s23}
t = - 2 \, a^{2} \, \frac{\mu}{L} \, \sqrt{1 - \epsilon^{2}} \, \tan^{-1} \left [  \frac{\epsilon 
- \tan \frac{\psi}{2}}{\sqrt{1 - \epsilon^{2}}}   \right ]
+ a^{2} \, \frac{\mu}{L} \, \frac{\epsilon \, (1 - \epsilon^{2}) \, \cos \psi}{\epsilon \, \sin \psi - 1}
+ \text{const} 
\end{equation}
For $\epsilon= 0$ we have the result
\begin{equation} \label{eq:w1}
t    =   a^{2} \, \frac{\mu}{L} \, \psi; \qquad \text{and} \qquad \frac{L}{a^{2} \, \mu}  = \omega_{y}
\end{equation}
The index $y$ stands for year. \\
 \\
From   (\ref{eq:s23}) we get the annual period
\begin{equation} \label{eq:s199}
\tau_{y} = t (\psi + 2 \, \pi) - t (\psi) = 2 \, \pi \, a^{2} \, \frac{\mu}{L} \,  \sqrt{1 - \epsilon^{2}}
 = 2 \, \pi \, \sqrt{\frac{a^{3}}{\gamma \, (M_{s} + m_{p})}}
\end{equation}
Using the time-angle equation   (\ref{eq:s23}) we can calculate the time intervals separating the typical values labelled solstices and equinoxes. Notice that, obviously, for the circular orbit we have (\ref{eq:w1})
 \begin{equation*} 
t    =   a^{2} \, \frac{\mu}{L} \, \psi; \qquad \text{or} \qquad \psi  = \omega_{y} \, t
\end{equation*}
Moreover the relationship time-angle is linear, therefore the solstices and equinoxes are equidistant in time. In fig. 
\ref{fig:gal4} we show what happens when $\epsilon = 0.5$, both for $\psi_{0} = 0$ and $\psi_{0} = \frac{\pi}{2}$

The time-angle equation (\ref{eq:s23}) is important (see appendix B). We show in fig. \ref{fig:gal5} the relationship $\psi, \, t$ in the case of $\epsilon = 0.017$ (this is the terrestrial orbit almost circular) and the case of
$\epsilon = 0.5$. \\
\begin{figure}[ptbh]
\begin{center}
\includegraphics[width=15 cm]{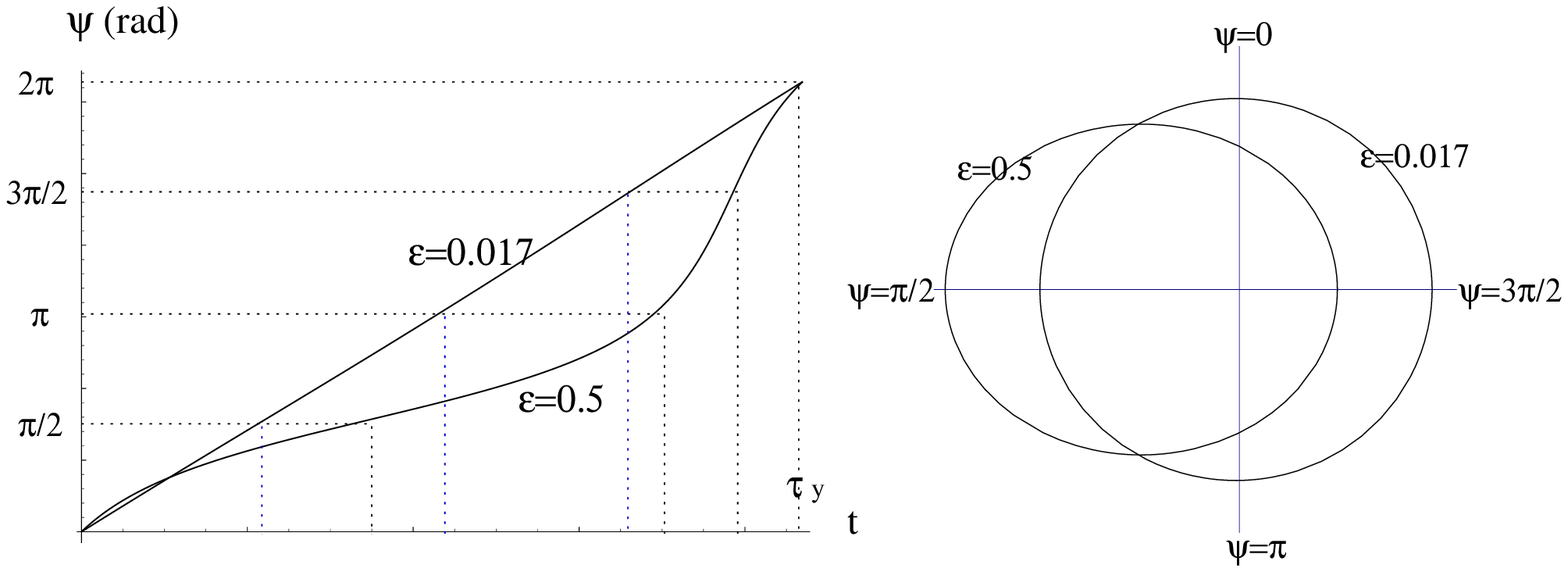} 
\caption{ \emph{ {\footnotesize  
 }}}  \label{fig:gal5}
\end{center}
\end{figure}
A point $Q$ on the surface of the planet with polar coordinates 
$R, \, \theta, \, \varphi$, has a complicated motion in the reference system of the star. We give a simplified example.

\begin{figure}[ptbh]
\begin{center}
\includegraphics[width=7 cm]{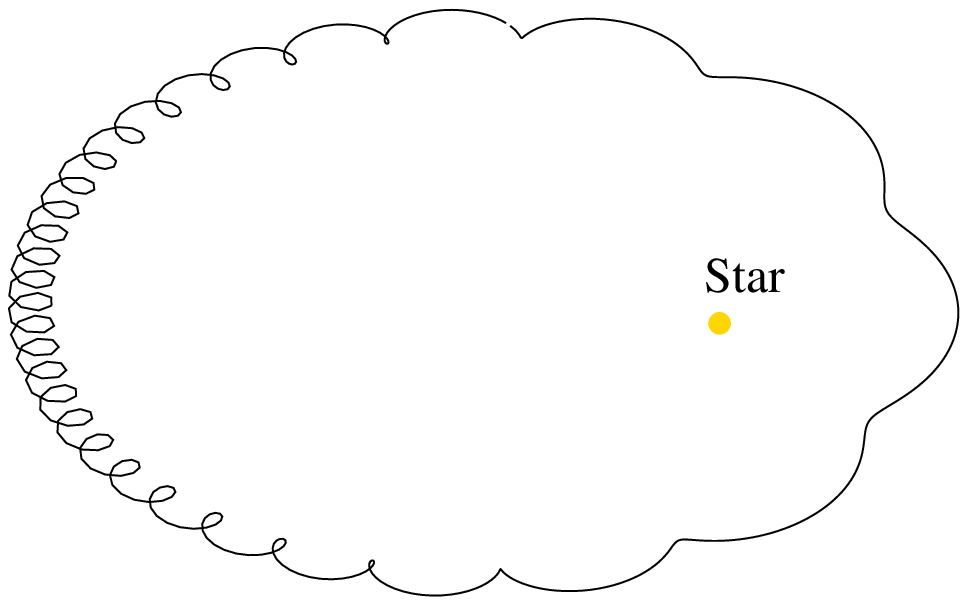}  $n_{1} \, \omega_{d} = 1 \, \omega_{y}$ \qquad \qquad \qquad \qquad \\
\includegraphics[width=7 cm]{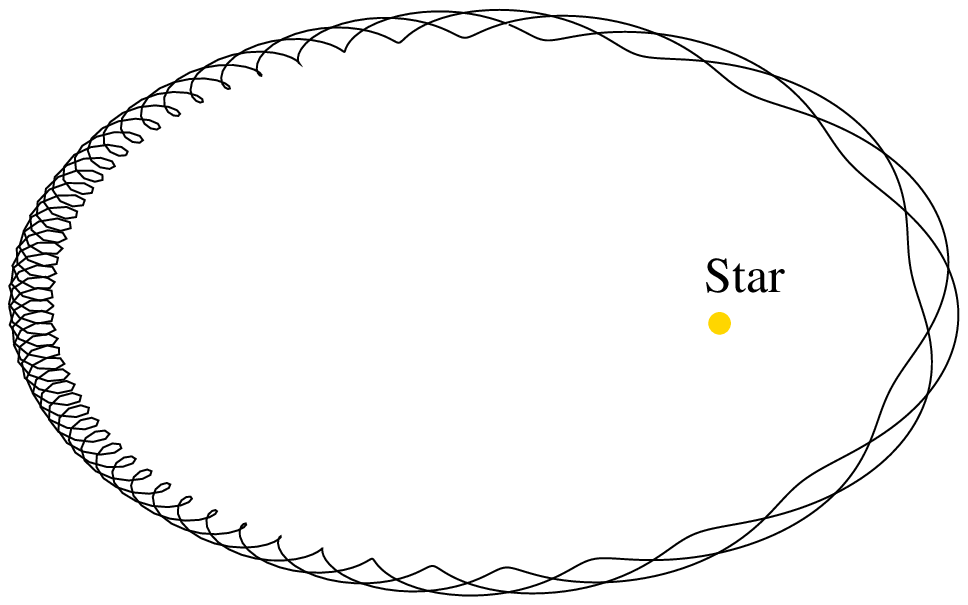} 
$n_{1} \, \omega_{d} = n_{2} \, \omega_{y}$ \qquad \qquad \qquad \qquad \\
\includegraphics[width=7 cm]{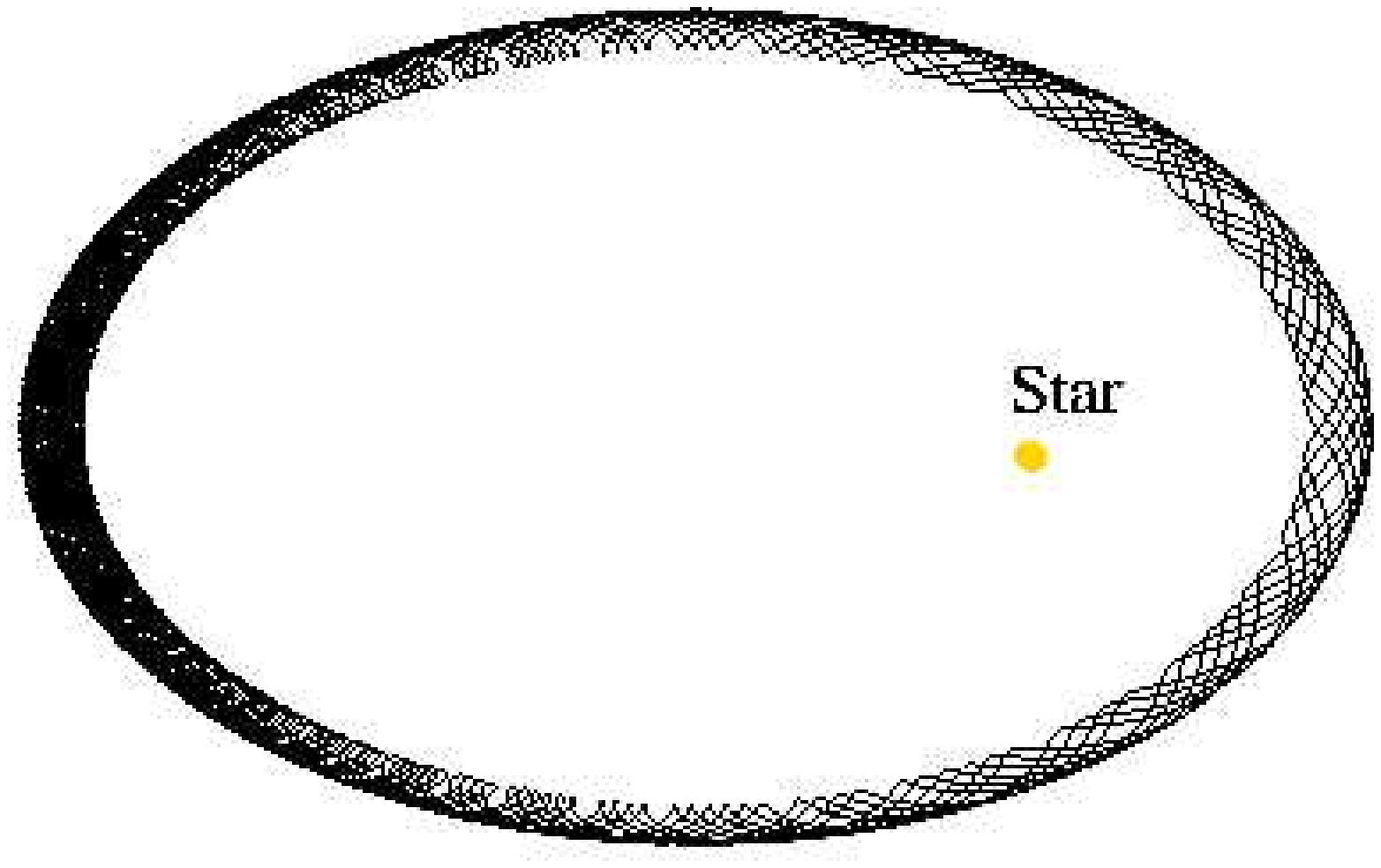} 
$ \omega_{d} = \text{real number} \cdot \omega_{y}$ \\
\caption{ \emph{ { \footnotesize  Motion of a point Q on the planet surface. The point Q rotates around the spin axis located in P with angular velocity $\omega_{d}$. The point P moves on the ellipse with focus F, position of the star, with angular velocity not constant. The relationship $\psi \leftrightarrow t$ is given by (\ref{eq:s23}). In the figure we have chosen
$\epsilon = 0.5$.
  }}}  \label{fig:gx6}
\end{center}
\end{figure}

If the spin of the planet is a constant vector perpendicular to the ecliptic plane, the point $Q$, at fixed $\theta$, rotates around the spin axis with angular velocity $\omega_{d}$ ($d$ for day).
We can write 
\begin{equation*}
\varphi = \omega_{d} \, t
\end{equation*}
The ratio between $\omega_{d} $ and $\omega_{y} = \frac{2 \, \pi}{\tau_{y}} $ can be a rational number or a real number.
These cases are illustrated in \ref{fig:gx6}

\section*{Appendix B. The incoming flux $\phi^{in}$}
\renewcommand{\theequation}{B.\arabic{equation}}
\renewcommand{\thefigure}{B.\arabic{figure}}
\setcounter{figure}{0}
\setcounter{equation}{0}
\subsection*{The Planck diluted spectrum} 

\begin{figure}[ptbh]
\begin{center}
\includegraphics[width=12 cm]{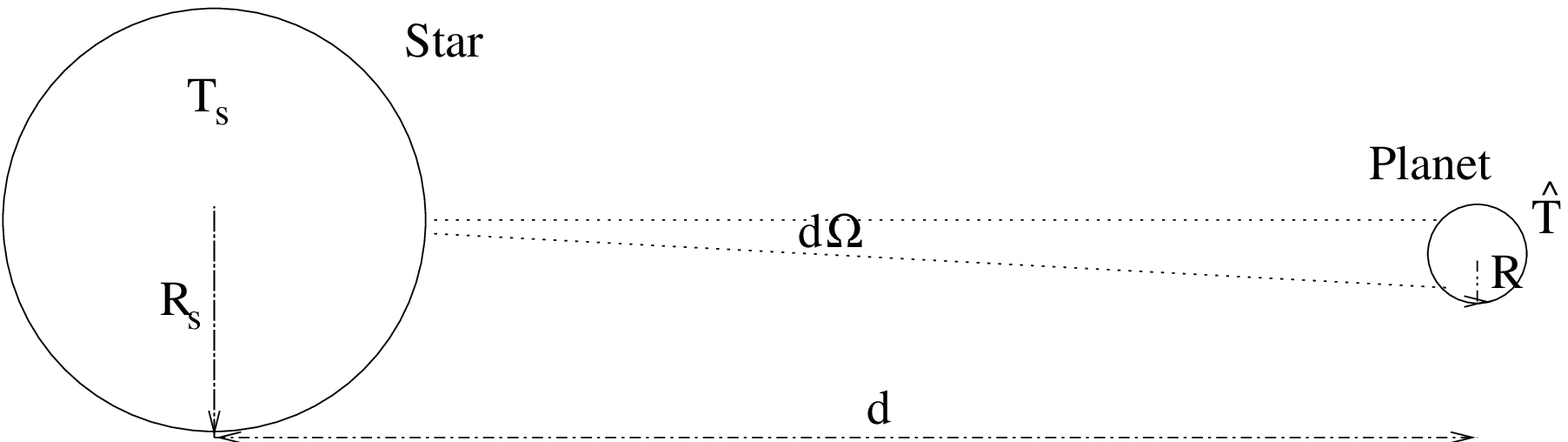}
\caption{} \label{fig:gal6}
\end{center}
\end{figure}

Let us discuss  first the concept of diluted Planck spectrum. We refer to fig. \ref{fig:gal6} \\
 \\ 
The conservation of energy in the solid angle $\ud \Omega$ gives us
\begin{equation} \label{eq:s2}
 R_{s}^{2} \, \sigma \, T_{s}^{4} =
 d^{2} \, \sigma \, \hat{T}^{4}
\end{equation}
$\hat{T}$ depends on  $T_{s}, \, R_{s}$ and $d$, the distance star-planet. Remember that for elliptic orbit $d =
d (\psi)$ and $\hat{T} = \hat{T} (\psi)$. 

$\hat{T}$ is the thermalization temperature, on a sphere of radius $d$, of the photons emitted by the star according to the Planck law
\begin{equation} \label{eq:su}
\mathcal{I}(T_{s},\nu) \, \ud \nu =   \frac{2 \pi \, h \, \nu^{3}}{c^{2}}
\frac{1}{e^{\frac{h \, \nu}{k \, T_{s}}} - 1}   \, \ud \nu; \qquad
[\mathcal{I} \, \ud \nu] = \frac{\text{W}}{\text{m}^{2}}
\end{equation}
 $h$ is the Planck constant, $c$ is the speed of light and $k$ is the Boltzmann constant.
The spectral distribution (\ref{eq:su}) at a distance $d$ is diluted by a factor  
 $\mathcal{D}_{s}$:
\begin{equation} \label{eq:s4}
\mathcal{D}_{s} = \frac{R_{s}^{2}}{d^{2}} =
 \frac{\hat{T}^{4}}{T_{s}^{4}}
\end{equation}
In other words at the distance $d$ arrives the diluted spectrum:
\begin{equation} \label{eq:s5}
\mathcal{I}_{dil}(T_{s},\nu) = \mathcal{D}_{s} \cdot \mathcal{I}(T_{s},\nu)
\end{equation}
The black surface positioned at $d$ has temperature $\hat{T}$ and in its turn emits with the spectrum
\begin{equation} \label{eq:s6}
\mathcal{I}(\hat{T},\nu) \, \ud \nu =
\frac{2 \pi \, h \, \nu^{3}}{c^{2}} \,
\frac{1}{\text{e}^{\frac{h \, \nu}{k \, \hat{T}} } - 1} \, \ud \nu
\end{equation}
\subsection*{The dependence of $\phi^{in}$ on $\psi, \, \theta, \, \varphi; \, \zeta$}

\begin{figure}[ptbh]
\begin{center}
\includegraphics[width=7 cm]{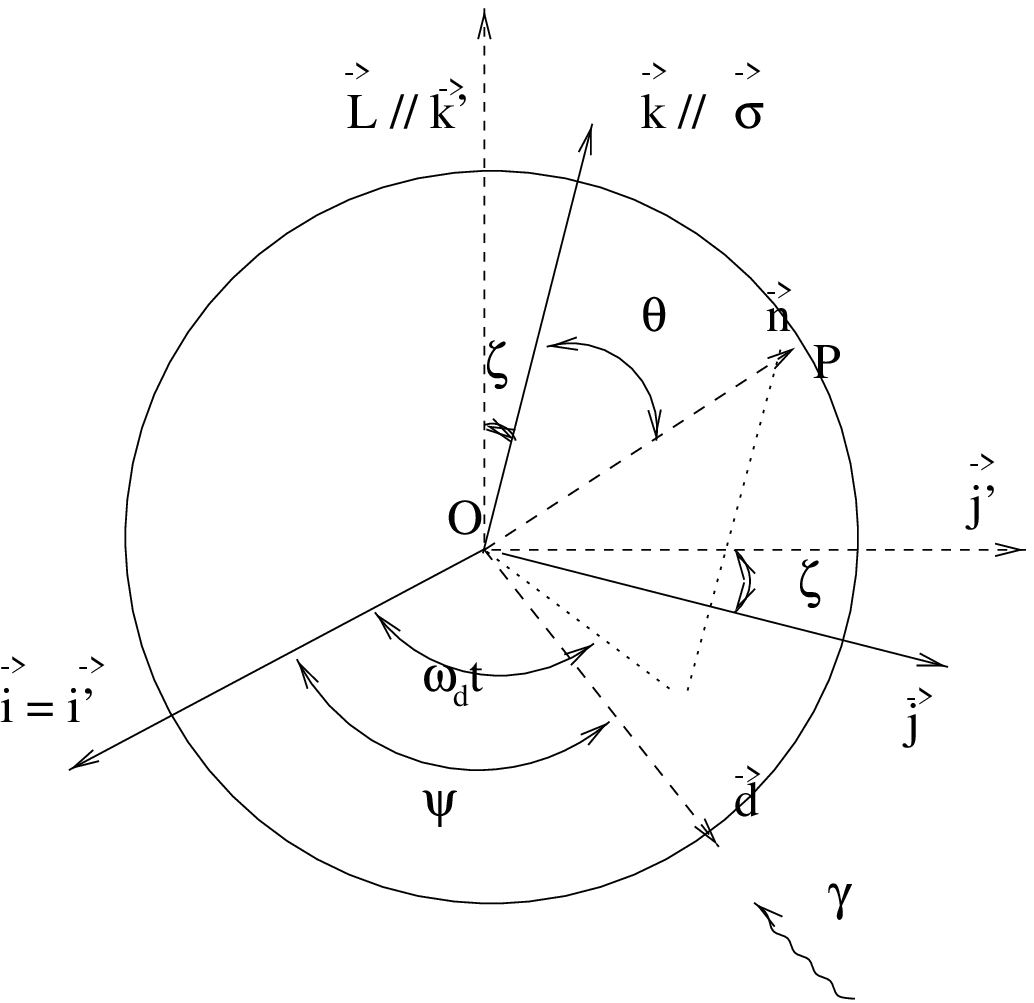}
\caption{ \emph{ \small {}}} \label{fig:spin}
\end{center}
\end{figure}

We can write the incoming photon flux $\phi^{in}$ for a point P on the surface of the planet with polar coordinates 
$R, \, \theta, \, \varphi$, where: \\
$R$ \, radius of the planet, \\
$0 \le \theta \le \pi$ is evaluated starting from the north pole ($\theta = 0$), and ending at the south pole $\theta = \pi$. \\
$0 \le \varphi \le 2 \, \pi$ \\
The point P, at fixed $\theta$, rotates around the spin axis with angular velocity $\omega_{d}$ ($d$ for day).
We can write 
\begin{equation*}
\varphi = \omega_{d} \, t
\end{equation*}
and is important to establish the relationship between $\varphi = 0, \, t = 0, \, \psi = 0$.
We chose  
\begin{equation} \label{eq:ca}
\varphi = 0, \, t = 0, \, \psi = 0 \qquad \qquad   \text{to be noon at spring equinox}
\end{equation}
With this convention we have 
\begin{equation} \label{eq:flux}
\phi^{in} (\psi, \theta, \varphi, t) = 
\sigma \, T_{s}^{4} \, \frac{R_{s}^{2}}{d^{2} (\psi)} \,\, \vec{d} \cdot \vec{n} \, \Theta (\vec{d} \cdot \vec{n})
= K (\psi) \, \mathfrak{f} (\theta, \varphi, \psi)
\end{equation}
using (\ref{eq:s20})
\begin{equation} \label{eq:kpsi} 
K (\psi) = \sigma \, T_{s}^{4} \, \frac{R_{s}^{2}}{d^{2} (\psi)} =
\sigma \, T_{s}^{4} \, \frac{R_{s}^{2}}{a^{2} \, (1 - \epsilon^{2})^{2}} \,
 (1 - \epsilon \, \sin \psi)^{2}
\end{equation} 
notice that  $\text{dim} \, (\phi^{in}) = \frac{\text{W}}{\text{m}^{2}}$.
\\
$\phi^{in} \, \ud S$ is the flux (in W) impinging on the surface element $\ud S$ positioned at P and having orientation $\vec{n}$ as shown in fig. \ref{fig:spin} \\

To understand eq. (\ref{eq:flux}), consider fig. \ref{fig:gal6} and \ref{fig:spin} jointly. The density of flux $\phi^{in}$ is the radiation seen from the planet, where it is the star that rotates according to the angle $\psi$,
(fig. \ref{fig:spin}). The unit vector $\vec{n}$ is the direction of OP, where OP has coordinates $R, \, 
\theta, \, \varphi$. The unit vector $\vec{d}$ is directed from the center of the planet O to the focus F (center of the star), 
fig. \ref{fig:gal3} and \ref{fig:gal6}.

\emph{Notice that in fig. \ref{fig:spin} the center of the planet is called O, while in fig. \ref{fig:gal3} the center 
of the planet is called P. }

The unit vector $\vec{d}$ has direction opposite to the momentum of the incoming photons
\begin{equation*} 
\vec{d} = - \frac{\vec{p}_{\gamma}}{|\vec{p}_{\gamma}|} 
\end{equation*}
After these remarks on the notation, by consideration of fig. \ref{fig:spin} we obtain
\begin{equation} \label{eq:for}
\begin{split}
 \vec{d} \cdot \vec{n} = &
\cos \theta \cdot \sin (\psi - \psi_{0} ) \cdot \sin \zeta + 
\sin \theta \cdot \cos (\psi - \psi_{0} ) \cdot \cos (\varphi_{0} + \omega_{d} \, t) + \\
& + \sin \theta \cdot \sin (\psi - \psi_{0} ) \cdot \cos \zeta  \cdot \sin (\varphi_{0} + \omega_{d} \, t)
\end{split}
\end{equation}
$\psi_{0}$ is the precession. In the following we put $\psi_{0} = 0$.
$\varphi_{0}$ is a given meridian. According to the origin of time (\ref{eq:ca}) we put $\varphi_{0} = 0$. There is no special identification for a meridian.

The ecliptic is the plane $\vec{i} ', \, \vec{j} '$ with normal unit vector $\vec{k} '$. The obliquity $\zeta$
is the angle $\vec{k}, \, \vec{k} '$. $\vec{d}$ rotates in the plane $\vec{i} ', \, \vec{j} '$.
The reference system of the planet is $\vec{i}, \, \vec{j}, \, \vec{k}$
The unit vector $\vec{n}$ rotates around $\vec{k}$.
The unit vector $\vec{d}$ is first rotated from the initial position $\vec{d} = \vec{i} '$ to the current position
$\vec{d} (\psi)$ with a rotation $R_{k'}$. The unit vector $\vec{n}$ is carried to the reference system $\vec{i}', \, \vec{j}', \,
\vec{k}'$ with a rotation $R_{\zeta}$. From these observations follows that the scale-product $\vec{d} \cdot \vec{n}$ is given explicitly by
\begin{equation*}
R_{\zeta}  = \left (
\begin{array}{ccc}
1 & 0 & 0 \\
 0 & \cos \zeta & \sin \zeta \\
 0 & - \sin \zeta & \cos \zeta \\
\end{array} \right ) \qquad \qquad \qquad 
R_{k'} = \left (
\begin{array}{ccc}
 \cos \psi & \sin \psi & 0 \\
 - \sin \psi & \cos \psi & 0 \\
0 & 0 & 1 \\
\end{array} \right )
\end{equation*}
\begin{equation*} 
 \vec{d} \cdot \vec{n} = \left (
\begin{array}{ccc}
1 &
0 &
0 \\
\end{array} \right ) \cdot \left [ \left (
\begin{array}{ccc}
 \cos \psi & \sin \psi & 0 \\
 - \sin \psi & \cos \psi & 0 \\
0 & 0 & 1 \\
\end{array} \right )     
\left (
\begin{array}{ccc}
1 & 0 & 0 \\
 0 & \cos \zeta & \sin \zeta \\
 0 & - \sin \zeta & \cos \zeta \\
\end{array} \right )  \left (
\begin{array}{c}
\sin \theta \, \cos \varphi \\
 \sin \theta \, \sin \varphi \\
\cos \theta \\
\end{array} \right )  \right ]
\end{equation*}
which is (\ref{eq:for}).

$\Theta (x)$ is the step function:
\begin{equation*}
\Theta (x) = \left \{ \begin{array}{ll}
1; & x > 0\\
0; & x < 0 \\
\end{array} \right .   
\end{equation*}
The expression $\vec{d} \cdot \vec{n}$ is a function of three variables $\psi, \, \theta, \, \varphi$, therefore
the constraint 
\begin{equation*}
 \vec{d} \cdot \vec{n}  > 0
\end{equation*}
determines a three dimensional domain in the space $\psi, \, \theta, \, \varphi$.

\begin{figure}
\hspace{-1cm}\includegraphics[width=18 cm]{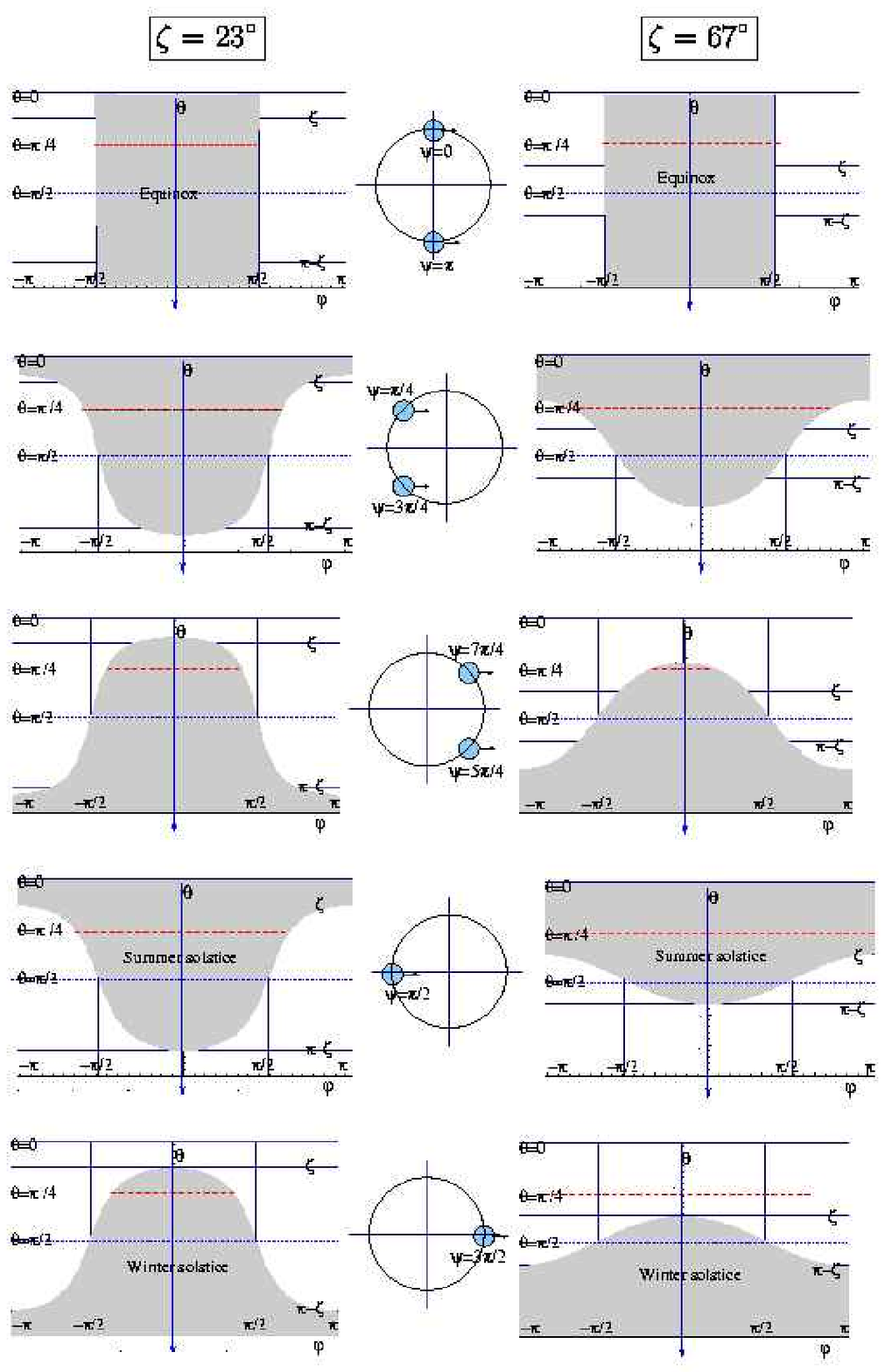}
\vspace{-5cm}
  \caption{ \emph{ {\footnotesize 
   }}}  \label{fig:gx7}
\end{figure}

We show in fig. \ref{fig:gx7} the domain $(\theta, \varphi)$ for various values of $\psi$. In these figures 
the precession is set to the value $\psi_{0} = 0$. The parameter $\zeta$ affects the domain  $\vec{d} \cdot \vec{n}  > 0$ in a very interesting way: we show the dependence on $\zeta$ presenting two columns, on the left $\zeta = 23^{\circ}$,
on the right $\zeta = 67^{\circ}$.

\subsection*{Harmonic expansion of $\phi^{in}$ } 
Remember the time angle relationship (\ref{eq:s23})
\begin{equation*}
 t = F (\psi); \qquad \qquad \psi = F^{-1} (t)
\end{equation*}
$\phi^{in}$ can be written as
\begin{equation} \label{eq:help3}
\phi^{in} = K (\psi) \, \mathfrak{f} (\theta, \varphi, \psi) =
\sigma \, T_{s}^{4} \, \frac{R_{s}^{2}}{a^{2} \, (1 - \epsilon^{2})^{2}} \,
 (1 - \epsilon \, \sin \psi )^{2} \, \mathfrak{f} (\theta, \varphi, \psi) 
\end{equation}
with $\varphi = \omega_{d} \, F (\psi)$. Alternatively $\phi^{in}$ can be written as
\begin{equation} \label{eq:help33}
\phi^{in} = K (t) \, \mathfrak{f} (\theta, \varphi, t) =
\sigma \, T_{s}^{4} \, \frac{R_{s}^{2}}{a^{2} \, (1 - \epsilon^{2})^{2}} \,
 (1 - \epsilon \, \sin \psi (t))^{2} \, \mathfrak{f} (\theta, \varphi, t) 
\end{equation}
with $\psi = F^{-1} (t)$, $\varphi = \omega_{d} \, t$. The expression (\ref{eq:help3}) is more natural considering the explicit expression of $\vec{d} \cdot \vec{n}$ and the fact that $t = F (\psi)$ is given explicitly in (\ref{eq:s23}).
Vice versa $\psi = F^{-1} (t)$ is given implicitly, or numerically.

We have seen that
$K (\psi)$ is periodic in the time-angle $\psi$ because $\frac{1}{d^{2} (\psi)}$ is periodic.
The factor $\mathfrak{f} (\theta, \varphi, \psi)$ requires a special discussion.
Consider eq. (\ref{eq:for}). We see that $\psi$ enters in terms $\sin \psi$ and $\cos \psi$, which are
periodic $2 \, \pi$, and enters also in the $\varphi$ dependence, which is of the kind $\sin \varphi$ and 
$\cos \varphi$. 

Since $\varphi = \omega_{d} \, t$ we write
\begin{equation*}
 \varphi = \omega_{d} \, F (\psi)
\end{equation*}
The question is whether $\cos (\omega_{d} \, F (\psi))$ or $\sin (\omega_{d} \, F (\psi))$
are periodic in $\psi$ or not. The answer is yes if $\omega_{d}  = n \, \omega_{y} $, in fact
in this case we write  
\begin{equation*}
  \omega_{d} \, t = n \, \omega_{y} \,  F (\psi) =  n \, \frac{2 \, \pi}{\tau_{y}} \,  F (\psi) 
\end{equation*}
but
\begin{equation*}
 \tau_{y} = F (2 \, \pi) - F (0)
\end{equation*}
and we can put $F (0) = 0$, using the integration constant in (\ref{eq:s23}), therefore
\begin{equation*}
  \omega_{d} \, t =   2 \, \pi  \, n \, \frac{F (\psi)}{F (2 \, \pi)} 
\end{equation*}
and consequently $\cos \left ( 2 \, \pi  \, n \, \frac{F (\psi)}{F (2 \, \pi)} \right )$ is periodic with period $n \, \tau_{y}$. On the contrary when $\omega_{d} = \alpha \, \omega_{y}$, with $\alpha$ real number
\begin{equation*}
  \omega_{d} \, t =   2 \, \pi  \, \alpha \, \frac{F (\psi)}{F (2 \, \pi)} 
\end{equation*}
In this case $\omega_{d} \, \tau_{y}$ is no longer a multiple of $2 \, \pi$ and in other words $\mathfrak{f} (\theta, \varphi, \psi)$ is no longer periodic in $\psi$.
This remark is very important for the study of the field $T (\vec{r}, t)$ which is driven by $\phi^{in}$.

In the following we discuss the case $\omega_{d} = n \, \omega_{y}$.

Let us consider the harmonic expansion in $\psi$, namely $\phi^{in}$ given by (\ref{eq:help3}).
The harmonic expansion of (\ref{eq:help3}) utilizes the orthonormal complete set
\begin{equation} 
W_{\ell m}^{M} (\theta, \varphi, \psi) = \frac{1}{\sqrt{2 \, \pi}} \,  Y_{\ell m} (\theta, \varphi) \cdot e^{i \, M \, \psi};
\end{equation}
\begin{equation} 
\int \, \ud \Gamma \, W_{\ell' m'}^{M' *} (\theta, \varphi, \psi) \, W_{\ell m}^{M} (\theta, \varphi, \psi) =
\delta_{\ell, \ell'} \, \delta_{m, m'} \, \delta_{M, M'}
\end{equation}
\begin{equation*} 
\int_{8 \, \pi^{2}} \, \ud \Gamma
=  \int_{0}^{2 \, \pi} \, \ud \psi \, \int_{0}^{2 \, \pi} \, \ud \varphi \, \int_{0}^{\pi} \, \sin \theta \, \ud \theta 
\end{equation*}
The expansion is
\begin{equation} \label{eq:zer}
\begin{split}
\phi^{in}  & =   
\sum_{\ell, m, M}  \, \phi_{\ell m M}  \cdot W_{\ell m}^{M} (\theta, \varphi, \psi);
\\ &  \ell   =  0, 1, 2, \ldots ;  
 \quad                  
m  =  - \ell  ,  - \ell   + 1, \ldots \ell; \quad M = 0, \pm 1, \pm 2, \ldots  \\
\end{split} 
\end{equation}                           
with
\begin{equation*} 
\phi_{\ell m M}  = \int_{8 \, \pi^2} \, \ud \Gamma \, \phi^{in} \,
 W_{\ell m}^{M *} (\theta, \varphi, \psi) 
\end{equation*}    
Notice that $W_{\ell m}^{M}$ are complex, $\phi_{\ell m M}$ are complex, but $\phi^{in}$ is real.
We know that
\begin{equation} \label{eq:help8}
<\phi^{in}>_{\theta, \varphi}  =
\frac{1}{4 \, \pi} \,
K (\psi)    \, \int \, \ud \Omega \, \mathfrak{f} (\theta, \varphi, \psi) = 
 \frac{1}{4 \, \pi} \,
K (\psi) \, \pi = \frac{1}{4} \,
K (\psi) 
\end{equation}
so that
\begin{equation} \label{eq:help80}
\frac{1}{\sqrt{2 \, \pi}} \, \sum_{M} \, e^{i \, M \, \psi} \, \sum_{\ell, m} \, <\phi^{in}_{\ell m M}  \,
Y_{\ell m} >_{\theta, \varphi}  =
 \frac{1}{4} \,
K (\psi) 
\end{equation}
but
\begin{equation*}
 \frac{1}{4 \, \pi} \,  \sum_{\ell, m} \, \int \, \ud \Omega \,
Y_{\ell m} =  Y_{00} =  \frac{1}{\sqrt{4 \, \pi}}
\end{equation*}
so that (\ref{eq:help80}) becomes
\begin{equation} \label{eq:help81}
\frac{1}{\sqrt{8 \, \pi^{2}}} \, \sum_{M} \, e^{i \, M \, \psi} \,  \phi^{in}_{00M}   = 
 \frac{1}{4} \,
K (\psi) 
\end{equation}
Since $K (\psi)$ is known explicitly (eq. (\ref{eq:kpsi})), namely
\begin{equation} \label{eq:kpsi1} 
\begin{split}
K (\psi) & = 
\sigma \, T_{s}^{4} \, \frac{R_{s}^{2}}{a^{2} \, (1 - \epsilon^{2})^{2}} \,
\left (1 - 2 \, \epsilon \, \sin \psi + \epsilon^{2} \, \frac{1 - \cos 2 \psi}{2} \right ) = \\
& = \mathcal{D} (\epsilon) \, \left (1 - 2 \, \epsilon \, \sin \psi + \epsilon^{2} \, \frac{1 - \cos 2 \psi}{2} \right )  \\
\end{split}
\end{equation}
Comparing like terms in (\ref{eq:help81}) and (\ref{eq:kpsi1}), we find
\begin{equation} \label{eq:kpsi2} 
\begin{array}{lll}
\phi^{in}_{00-2} =  \displaystyle{-   \frac{\pi}{4 \, \sqrt{2}} \, \epsilon^{2} \, \mathcal{D} (\epsilon); }  &  &  \phi^{in}_{002} = 
 \displaystyle{-   \frac{\pi}{4 \, \sqrt{2}} \, \epsilon^{2} \, \mathcal{D} (\epsilon); } \\ \\
\phi^{in}_{00-1} =  \displaystyle{ -  i \, \frac{\pi}{\sqrt{2}} \,  \epsilon \, \mathcal{D} (\epsilon); }   &  &  \phi^{in}_{001} = 
 \displaystyle{ i \, \frac{\pi}{\sqrt{2}} \, \epsilon \, \mathcal{D} (\epsilon); }
 \\ \\
\phi^{in}_{000} =  \displaystyle{\frac{\pi}{2 \, \sqrt{2}} } \, (2 + \epsilon^{2}) \, \mathcal{D} (\epsilon) &      \\
\end{array}
\end{equation}
Obviously the component $_{000}$ is the largest and the only one that survives in the $\epsilon \to 0$ limit.
The annual average of the incoming flux is
 \begin{equation} \label{eq:help88}
<\phi^{in}>_{\theta, \varphi, \psi}  =
\frac{1}{2 \, \pi} \, \frac{1}{4} \,
 \int \, \ud \psi \,  K (\psi) =  \frac{1}{8} \, (2 + \epsilon^{2}) \, \sigma \, T_{s}^{4} \, \frac{R_{s}^{2}}{a^{2} \, (1 - \epsilon^{2})^{2}}
\end{equation}
Next let us consider the harmonic expansion in $t$, namely $\phi^{in}$ given by (\ref{eq:help33}).
The factor $K (t)$ is periodic in $t$ with period $\tau_{y}$. The orthonormal set is
\begin{equation} 
W_{\ell m M} (\theta, \varphi, t) = \frac{1}{\sqrt{\tau_{y}}} \,  Y_{\ell m} (\theta, \varphi) \cdot e^{i \, \frac{2 \, \pi}{\tau_{y}} \, M \, t};
\end{equation}
The expansion is
\begin{equation} \label{eq:zer00}
\begin{split}
\phi^{in}  & =   
\sum_{\ell, m, M}  \, \phi_{\ell m M}^{in}  \cdot W_{\ell m}^{M} (\theta, \varphi, t);
\\ &  \ell   =  0, 1, 2, \ldots ;  
 \quad                  
m  =  - \ell  ,  - \ell   + 1, \ldots \ell; \quad M = 0, \pm 1, \pm 2, \ldots  \\
\end{split} 
\end{equation}                           
with
\begin{equation*} 
\phi_{\ell m M}^{in}  = \int_{\tau_{y}} \ud t \, \int \ud \Omega \, \phi^{in} \,
 W_{\ell, m}^{M *} (\theta, \varphi, t) 
\end{equation*}    
Let us perform the average over $\theta, \varphi$
\begin{equation} \label{eq:help800}
\begin{split}
<\phi^{in}>_{\theta, \varphi}  & =
\frac{1}{4 \, \pi} \, \frac{1}{\sqrt{\tau_{y}}} \,
  \sum_{\ell m M} \,  \int \, \ud \Omega \, \phi_{\ell m M}^{in} \,  Y_{\ell m} \, e^{i \, \frac{2 \, \pi}{\tau_{y}} \, M \, t} = \\
& =   \frac{1}{\sqrt{\tau_{y}}} \,
  \sum_{M} \,   \phi_{0 0 M}^{in} \,  Y_{0 0} \, e^{i \, \frac{2 \, \pi}{\tau_{y}} \, M \, t} 
\end{split}
\end{equation}
But we know that
\begin{equation} \label{eq:cas}
<\phi^{in}>_{\theta, \varphi}  = \frac{1}{4} \, K (t)
= \frac{1}{4} \, \frac{1}{\sqrt{\tau_{y}}} \, \sum_{M} \,  K_{M}  \, e^{i \, \frac{2 \, \pi}{\tau_{y}} \, M \, t} 
\end{equation}
The expansion (\ref{eq:help800}) contains all the Fourier components, $M = 0, \, \pm 1, \, \pm 2 \ldots$
and is not restricted to five Fourier components as it is with the expansion in $\psi$, formula (\ref{eq:kpsi2}).

Concerning the non averaged expansion (\ref{eq:zer00}) in the case we are considering, namely $\omega_{d} = n \, \omega_{y}$
or $\tau_{y} = n \, \tau_{d}$ we must have a component $M = n$ that is the second large mode of the expansion.
Nevertheless eq. (\ref{eq:zer00}) contains all the other modes. 

It is clear that the expansion in $\psi$ is helpful in order to appreciate the kinematics of the elliptic orbit; the expansion in $t$ is necessary for the study of the Fourier heat equation given in appendix C.

Finally we consider the case $\frac{\omega_{d}}{\omega_{y}} = $real number.
In this case the domain $\theta, \, \varphi$ is always bounded by $4 \, \pi$, but the domain $t$ is unbounded.
It follows that the ortho-normal set corresponds to discrete indices with respect to $\theta, \, \varphi$  and a continuous index
with respect to the time $t$.

The eigenfunctions are
\begin{equation} 
W_{\ell m}^{\nu}  = \frac{1}{\sqrt{2 \, \pi}} \,  Y_{\ell m} (\theta, \varphi) \cdot e^{i \, \nu \,  t}
\end{equation}
and the orthonormality relations 
\begin{equation} \label{eq:ort1}
 \frac{1}{2 \, \pi} \, \int_{4 \, \pi} \, \ud \Omega \, \int_{0}^{\infty} \, \ud t \,
 Y_{\ell' m'}^{*}  \,Y_{\ell m}  \, e^{i \, (\nu - \nu') \, t} =
\delta_{\ell, \ell'} \, \delta_{m, m'} \, \delta (\nu - \nu')
\end{equation}
The harmonic expansion is
\begin{equation} 
\begin{split}
\phi^{in}  & =   
\sum_{\ell, m}  \, \int  \ud \nu \, \phi_{\ell m \nu}^{in}  \cdot W_{\ell m}^{\nu} (\theta, \varphi, t);
\\ &  \ell   =  0, 1, 2, \ldots ;  
 \quad                  
m  =  - \ell  ,  - \ell   + 1, \ldots \ell;   \\
\end{split} 
\end{equation}                           
with
\begin{equation*} 
\phi_{\ell m \nu}^{in}  = \int_{0}^{\infty} \ud t \, \int \, \ud \Omega \, \phi^{in} \,
 W_{\ell, m}^{\nu *} (\theta, \varphi, t) 
\end{equation*}   
The interesting fact is that $\phi^{in}$ is non periodic in $t$ but its $\theta, \, \varphi$ average is periodic in $t$.
\begin{equation} \label{eq:cas1}
<\phi^{in}>_{\theta, \varphi}  = \frac{1}{4} \, \frac{1}{\sqrt{\tau_{y}}} \, \sum_{M} \, K_{M}  \, e^{i \, \frac{2 \, \pi}{\tau_{y}} \, M \, t} 
\end{equation}
as we have seen above in formula (\ref{eq:cas}).

Dimensional remarks for the expansion in $t$.
\begin{equation*} 
\begin{array}{l}
\displaystyle{\text{dim} \, (\phi^{in}) = \frac{\text{W}}{\text{m}^{2}} } \\
\\
\displaystyle{\text{dim} \, (W_{\ell, m}^{M} (\theta, \varphi, t) ) = \text{s}^{-\frac{1}{2}} } \\ \\
\displaystyle{\text{dim} \, (W_{\ell, m}^{\nu} (\theta, \varphi, t) ) = \text{dimensionless} } \\ \\
\displaystyle{\text{dim} \, (\phi_{\ell m M}^{in}) = \frac{\text{W}}{\text{m}^{2}} \, \text{s}^{\frac{1}{2}} } \\ \\
\displaystyle{\text{dim} \, (\phi_{\ell m \nu}^{in}) = \frac{\text{W}}{\text{m}^{2}} \cdot \text{s} } \\ \\
\displaystyle{\text{dim} \, (\delta (\nu)) = \text{s}} \\
\end{array}
\end{equation*}



\section*{Appendix C. The temperature field.}
\renewcommand{\theequation}{C.\arabic{equation}}
\renewcommand{\thefigure}{C.\arabic{figure}}
\setcounter{figure}{0}
\setcounter{equation}{0}
The study of the temperature field develops along these steps.
\begin{enumerate}
\item The planet is assumed to be homogeneous, namely $\rho_{v}$ density, $c_{v}$, specific heat, $\kappa$, Fourier coefficient
are constant in the crust. The planet is made of a hot core and a crust according to this figure. \\
\begin{figure}[ptbh]
\begin{center}
\includegraphics[width=5 cm]{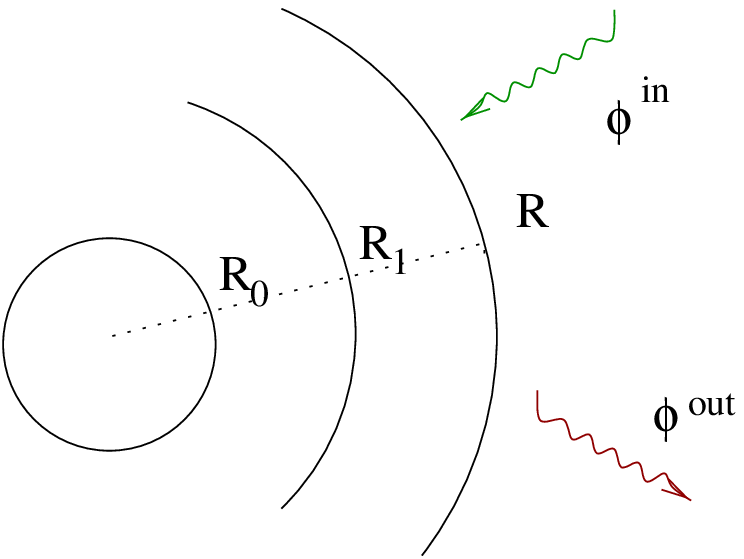} 
\caption{ \emph{ { \small 
} }}
\label{fig:ter}
\end{center}
\end{figure}

For $r < R_{0}$ there is a hot core which cools off and produces the source for the
endogenous temperature field decreasing from $R_{0}$
to $R$. 

From the star comes the external  input $\phi^{in}$. For $r < R_{1}$ the temperature flow generated by the input $\phi^{in}$ is
negligible.
 With the assumption of homogeneity the Fourier equation for  the temperature field $T (\vec{r}, t)$
is conveniently written in polar coordinates
\begin{equation} \label{eq:a1}
c_{v} \, \rho_{v} \, \frac{\de T}{\de t} = 
  \kappa \,  \Delta_{r, \theta, \varphi} \, T; \qquad \qquad R_{0} < r < R
\end{equation}
\begin{equation*} 
\Delta_{r \theta \varphi}  =  \frac{\de^{2}}{\de r^{2}} +  \frac{2}{r} \,  \frac{\de}{\de r} +           
 \frac{1}{r^{2}} \, \left \{ \frac{\de^{2} }{\de \theta^{2}} + 
\frac{1}{\tan \theta}  \, \frac{\de }{\de \theta}   +   \frac{1}{\sin^{2} \theta}
\frac{\de^{2}}{\de \varphi^{2}} \right \}
\end{equation*}
Dimensionally eq. (\ref{eq:a1}) is  $\frac{\textrm{W}}{\textrm{m}^{3}} $.
As equation (\ref{eq:a1})  is linear, the general solution is the sum of the endogenous solution belonging to the boundary condition at $r = R_{0}$
\begin{equation} 
T (R_{0}, t) = T_{core} (t)
\end{equation}
the cooling of the core, plus the solution driven by the incoming flux $\phi^{in} (t)$. Both $\phi^{in} (t)$ and $T_{core} (t)$ 
are inputs independent of each other. We write
\begin{equation} 
T_{gen} (\vec{r},t) = T_{core} (t) + T (\vec{r},t)
\end{equation}
The behaviour of the temperature field from $ T_{core} (t)$ has been studied in a a preceding paper \cite{ref:ava}.
Here we concentrate on the star driven temperature field.

The statement of the problem implies
\begin{description}
\item[-]  The differential equation (\ref{eq:a1})
\item[-]   The imposed boundary condition, or driving term
\item[-] The initial  condition
\end{description}
We can take into account the driving term writing
\begin{equation} \label{eq:c1}
c_{v} \, \rho_{v} \, \frac{\de T}{\de t} = 
  \kappa \,  \Delta_{r, \theta, \varphi} \, T + \phi^{in} (\theta, \varphi, t) \, \delta (r - R)
\end{equation}
with $\delta (x)$ Dirac delta function.
\item Let us comment the dissipative nature of equation (\ref{eq:c1}). Consider a shell $r - \tilde{\delta} < r < R$.
(in the following we will justify the concept of shell $\tilde{\delta}$) and take uniform temperature within this shell
\begin{equation*} 
T (R - \tilde{\delta}, \theta, \varphi, t) \sim T (R, \theta, \varphi, t) = T (t)
\end{equation*}
with initial condition
\begin{equation} 
T (t = 0) = T_{0}
\end{equation}
Correspondingly, consider the total flux  of $\Phi^{in}_{s}$. The index $s$ stands for surface.
Such quantity has been discussed in appendix B.
\begin{equation} 
\Phi^{in}_{s} = \pi \, K (t) \, R^{2}
\end{equation}
Now we integrate equation (\ref{eq:c1}) in the volume
\begin{equation} 
V = 4 \, \pi \, (R - \tilde{\delta})^{2} \, \tilde{\delta} 
\end{equation}
having external surface $\mathcal{S}$ and internal surface $\mathcal{S}_{i}$.
We get
\begin{equation} \label{eq:c2}
c_{v} \, \rho_{v} \, V \, \frac{\ud T}{\ud t} \simeq 
  \kappa \,  \int_{\mathcal{S}} \, \nabla T \cdot \vec{n} \, \ud \mathcal{S} + 
\Phi^{in}_{s} (t)
\end{equation}
(we have used the Gauss lemma relating volume integral to surface integral and we have neglected the contribution to the surface integral across the internal surface $\mathcal{S}_{i}$). Now the term
\begin{equation*} 
  \kappa \,  \int_{\mathcal{S}} \, \nabla T \cdot \vec{n} \, \ud \mathcal{S} 
\end{equation*}
is the outgoing flux. As the planet is positioned in the vacuum, this term is
\begin{equation} 
 - \kappa \,  \int_{\mathcal{S}} \, \nabla T \cdot \vec{n} \, \ud \mathcal{S} =
\Phi^{out} = 4 \, \pi \, R^{2} \, \sigma \, T^{4} (t)
\end{equation}
Putting the pieces together we have
\begin{equation} \label{eq:c3}
c_{v} \, \rho_{v} \, V \, \frac{\ud T}{\ud t} = 
 \pi \, R^{2} \, K (t)  -  4 \, \pi \, R^{2} \, \sigma \, T^{4} (t)
\end{equation}
The non linear equation (\ref{eq:c3}) has the periodic forcing term  $\pi \, K (t) \, R^{2}$. Let us consider
the period average of the forcing term
\begin{equation*} 
\frac{1}{\tau_{y}} \,  \int_{t}^{t + \tau_{y}} \, K (t') \, \ud t' = K_{0} = \text{constant}
\end{equation*}
and consider
\begin{equation*} 
\frac{1}{\tau_{y}} \,   \int_{t}^{t + \tau_{y}} \, T (t') \, \ud t' = \tilde{T} (t)
\end{equation*}
Finally consider the segmented equation \cite{ref:lot}
\begin{equation} \label{eq:c4}
c_{v} \, \rho_{v} \, V \, \frac{\ud \tilde{T}}{\ud t} \simeq 
 K_{0} \, \pi \,  R^{2} -  4 \, \pi \, R^{2} \, \sigma \, \tilde{T}^{4} (t)
\end{equation}
This equation has the asymptotic fixed point
\begin{equation} \label{eq:c5}
 \tilde{T}_{as} = 
\left ( \frac{K_{0}}{4 \, \sigma} \right )^{\frac{1}{4}}
\end{equation}
Linearizing around $\tilde{T}_{as}$, we write
\begin{equation} \label{eq:c6}
\tilde{T} = \tilde{T}_{as} + \xi
\end{equation}
\begin{equation*} 
  \tilde{T}^{4} = \tilde{T}_{as}^{4} + 4 \, \tilde{T}_{as}^{3} \, \xi
\end{equation*}
and we get the equation
\begin{equation} \label{eq:c50}
c_{v} \, \rho_{v} \, V \, \dot{\xi} = 
  -  16 \, \pi \, R^{2} \, \sigma \, \tilde{T}_{as}^{3} \, \xi
\end{equation}
which has solution
\begin{equation} \label{eq:c60}
\xi = \xi_{0} \,
 e^{ \lambda \, t}
\end{equation}
where
\begin{equation} \label{eq:c7}
 \lambda = - \frac{16 \, \pi \, R^{2} \, \sigma \, \tilde{T}_{as}^{3}}{c_{v} \, \rho_{v} \, V} ;
\qquad \qquad \text{dim} \, (\lambda) = \text{s}^{-1}
\end{equation}
is the Liapounov exponent of the stable fixed point $\tilde{T}_{as}$.

In conclusion $T (t)$ is asymptotically oscillating with period $\tau_{y}$.
We use the preceding analysis as a justification to restrict the study of
$T (r, \theta, \varphi, t)$ to the asymptotic regime, where there is no longer memory of the initial condition, because the field $T (r, \theta, \varphi, t)$ has settled down on the periodic attracting input term 
$\phi^{in} (\theta, \varphi, t)$; in other words we neglect the transient.
\item The harmonic expansion is a way to diagonalize the differential equation (\ref{eq:c1}).
Such program cannot be carried to the end because we do not have the explicit expression of $\phi^{in}_{\ell m M}$,
nevertheless from the harmonic expansion of $T$ we can get important informations.

Let us expand $T$ using the orthonormal complete set
\begin{equation} 
W_{\ell m M}  = \frac{1}{\sqrt{\tau_{y}}} \,  Y_{\ell m} (\theta, \varphi) \cdot e^{i \, M \, \frac{2 \, \pi}{\tau_{y}} \, t}
\end{equation}
with
\begin{equation} \label{eq:ort}
 \frac{1}{\tau_{y}} \, \int_{4 \, \pi} \, \ud \Omega \, \int_{\tau_{y}} \, \ud t \,
 Y_{\ell' m'}^{*}  \,Y_{\ell m}  \, e^{i \, (M - M') \, \frac{2 \, \pi}{\tau_{y}} \, t} =
\delta_{\ell, \ell'} \, \delta_{m, m'} \, \delta_{M, M'}
\end{equation}
The harmonic expansion is
\begin{equation} \label{eq:ser}
\begin{split}
T(r, \theta, \varphi, t)  & =   
\sum_{\ell, m, M} \,  \, T_{\ell m M} (r) \cdot W_{\ell m M}  = \\
& = \frac{1}{\sqrt{\tau_{y}}} \, \sum_{\ell, m M}  \, T_{\ell m M} (r) \cdot Y_{\ell m} (\theta, \varphi) \cdot 
e^{i \, M \, \frac{2 \, \pi}{\tau_{y}} \, t} 
\\ &  \ell   =  0, 1, 2, \ldots ;  
 \quad                  
m  =  - \ell  ,  - \ell   + 1, \ldots \ell;  \quad
 M   =  0, \pm 1, \pm 2, \ldots \\
\end{split} 
\end{equation}                           
where
\begin{equation} \label{eq:ser00}
T_{\ell m M} (r) = \frac{1}{\sqrt{\tau_{y}}} \, \int_{\tau_{y}} \, \ud t \,
   \int_{4 \, \pi} \, \ud \Omega \, T(r, \theta, \varphi, t) \,
 Y_{\ell, m}^{*} (\theta, \varphi) \,  	e^{- i \, M \, \frac{2 \, \pi}{\tau_{y}} \, t}
\end{equation}  
Inserting (\ref{eq:ser}) into (\ref{eq:a1}) we get
\begin{equation} \label{eq:lal}
    \sum_{\ell ,m, M}  \, W_{\ell m M}  \cdot  \left [  \frac{\de^{2}}{\de r^{2}}     +  \frac{2}{r} \, \frac{\de}{\de r} 
-  \left ( \frac{\ell  (\ell  + 1)}{r^{2}} + \frac{i \, c_{v} \, \rho_{v} \, M \, \omega_{y}}{\kappa} \right ) \right ]
\, T_{\ell m M} (r)  = 0; 
\end{equation}  
We project eq. (\ref{eq:lal})  on $W_{\ell' m' M'}^{*} $ and we integrate in $\ud \Omega \, \ud t$. Using the ortho-normality
(\ref{eq:ort}) we obtain
\begin{equation} \label{eq:bes}
 \frac{\ud^{2} T_{\ell m M} (r)}{\ud r^{2}}     +  \frac{2}{r} \, \frac{\ud T_{\ell m M} (r)}{\ud r} 
-  \left ( \frac{\ell  (\ell  + 1)}{r^{2}} + \frac{i \, c_{v} \, \rho_{v} \, M \, \omega_{y}}{\kappa} \right ) T_{\ell m M} (r)  = 0
\end{equation}  
\item The radial behaviour. \\
Equation (\ref{eq:bes})
is a particular case of the Tricomi equation \label{bat} \cite{ref:bat}. We are interested in the solution 
of (\ref{eq:bes}) which is regular at $ r = 0 \, $ and behaving as a positive exponential for large  $ r $; from
\cite{ref:bat} we get:
\begin{equation} \label{eq:asint}
T_{\ell m M} (r) \, _{\widetilde{_{r \rightarrow \infty}}} \, \, H_{\ell m M}  \, \, \frac{1}{r} \, \, 
e^{r \, \sqrt{\frac{c_{v} \, \rho_{v} \, \omega_{y}}{2 \, \kappa} \, M} + i \, r \, \sqrt{\frac{c_{v} \, \rho_{v} \,
\omega_{y}}{2 \, \kappa} \, M}}                                                                                            
\end{equation}
Equation (\ref{eq:asint}) contains a relevant physical information. In fact consider the mode $M = 1$ corresponding to the 
frequency $\frac{1}{\tau_{y}}$ and the mode $M = n$ corresponds to the frequency $\frac{1}{\tau_{d}} = \frac{n}{\tau_{y}}$.
Let us define ``penetration length'' for the mode $M$ the value $\tilde{\delta}_{M}$ such that
\begin{equation}
\text{exp} \left [
   \left ( (R - \tilde{\delta}_{M}) - R \right ) \cdot
\sqrt{\frac{c_{v} \, \rho_{v}}{2 \, \kappa} \, M \, \omega_{y}} \right ] =  \frac{1}{e}                                                                                      
\end{equation}
We have
\begin{equation} \label{eq:penet}
 \tilde{\delta}_{M}  =  \sqrt{\frac{2 \, \kappa}{c_{v} \, \rho_{v} \, M \, \omega_{y}}} 
\end{equation} 
So the various higher Fourier components generated by the input frequencies     
$\nu_{y} = \frac{1}{\tau_{y}}$ and $\nu_{d} = \frac{1}{\tau_{d}}$, with large $M$, have decreasing penetration. In 
other words, they affect a thin layer near the surface.
Eq. (\ref{eq:asint}) tells us which is the radial dependence of mode $M_{1} = 1$ and the radial dependence of the mode  
$M_{2} = n$. From (\ref{eq:penet})  we see that
\begin{equation} \label{eq:qrl} 
\frac{\tilde{\delta}_{M_{1}}}{\tilde{\delta}_{M_{2}}}  = \sqrt{\frac{M_{2}}{M_{1}}}                                                                              
\end{equation}
This square root law governs the penetration of the daily temperature oscillation with respect to the annual temperature oscillation.

Notice that when $\omega_{d} = \alpha \, \omega_{y}$, with $\alpha$ irrational number, the Fourier analysis of $T$
with respect to the time $t$ is continuous. No matter how complicated the spectral analysis of $T$ is \cite{ref:bpv},
namely the explicit dependence of $H_{\ell m \nu}$ on $\nu$, the square root law (\ref{eq:qrl}) remains valid.
\item The angular behaviour. \\
Note that the coefficients $H_{\ell m M}$ appear as a common factor to the addenda of equation (\ref{eq:bes}), and remain, at this point, undetermined. To find  $H_{\ell m M}$ we need to use the boundary condition, namely the equation
\begin{equation} \label{eq:gtr}
\left .  \kappa \,   \frac{\de T}{\de r} \right |_{r = R}  = - \sigma \, T^{4} (R) + \phi^{in}
\end{equation}
In fact inserting the expansion (\ref{eq:ser}) with the explicit solution (\ref{eq:asint}) evaluated at $r = R$ into (\ref{eq:gtr}),
we get an algebraic equation involving $H_{\ell m M}$ and $\phi^{in}_{\ell m M}$. In principle this equation gives the coefficients
$H_{\ell m M}$. Unfortunately there are two difficulties. 

The coefficients    are not known analytically, and this happens because $\mathfrak{f} (\theta, \varphi, t)$ is a discontinuous function. This function could be rounded off with some approximations.

The second difficulty is that the algebraic equation relating $H_{\ell m M}$ and $\phi^{in}_{\ell m M}$ contains the fourth power of a series. In this case is necessary to perform a truncation. We have studied both procedures, rounding off and truncation, in a preceding paper, concerning a particular case where the function $\mathfrak{f} (\theta, \varphi, t)$ was simpler.
Here we skip the analysis and we leave the study of the angular dependence of the field $T$ to the numerical calculation, appendix D.

\item Finally we consider the case $\frac{\omega_{d}}{\omega_{y}} = $real number.
In this case we adopt the same harmonic expansion that has been used in appendix B for the input function 
$\phi^{in} (\theta, \varphi, t) $.
\begin{equation} \label{eq:cc1}
T    = 
\sum_{\ell, m}  \, \int  \ud \nu \, T_{\ell m \nu}  \cdot W_{\ell m}^{\nu} (\theta, \varphi, t)
\end{equation}                           
with
\begin{equation} 
T_{\ell m \nu}  = \int_{0}^{\infty} \ud t \, \int \, \ud \Omega \, T \,
 W_{\ell, m}^{\nu *} (\theta, \varphi, t) 
\end{equation}   
The expansion (\ref{eq:cc1}) is unrestricted with respect to the time interval. As we are dealing with a dissipation problem (the Fourier heat equation) the time runs from $t = 0$ to $t \to \infty$ rather than $- \infty < t < \infty$ as we have for Hamiltonian problems.

On the other hand the solution $T$ of the equation (\ref{eq:a1}), as we have seen in (\ref{eq:c5}) is asymptotically attracted by the
average driving term in the sense that
\begin{equation} \label{eq:c500}
 <T>_{\theta, \varphi} \, _{\overrightarrow{t \to \infty}} \,  
\left ( \frac{1}{\sigma} \, <\phi^{in}>_{\theta, \varphi}   \right )^{\frac{1}{4}}
\end{equation}
This result tells us that in the asymptotic regime (\ref{eq:c500}) we can write
\begin{equation} \label{eq:c501}
\begin{split}
 <T>_{\theta, \varphi} & = \frac{1}{4 \, \pi} \, \int \ud \Omega \, T  = 
\frac{1}{\sqrt{8 \, \pi^{2}}} \, \int \ud \nu \, T_{00\nu}  \,  e^{i \, \nu \, t} \, _{\overrightarrow{t \to \infty}}
\\
& _{\overrightarrow{t \to \infty}} \,  \frac{1}{\sqrt{4 \, \pi \, \tau_{y}}} \, \sum_{M}  \, T_{00M}  \,  e^{i \, \frac{2 \, \pi}{\tau_{y}} \, M \, t} \\
\end{split}
\end{equation}
Equation (\ref{eq:c501}) contains the property that the solution $T$ of equation (\ref{eq:a1}) is attracted to the periodic behaviour 
(\ref{eq:c501}).

Another way to express the same result is to write the average of $T$ in the interval of time $\tau_{y}$:
\begin{equation} \label{eq:c502}
\begin{split}
\frac{1}{\tau_{y}} \, &  \int_{t}^{t + \tau_{y}} \ud t' \, <T>_{\theta, \varphi}   = 
\frac{1}{\tau_{y}} \, \frac{1}{\sqrt{8 \, \pi^{2}}} \,  \int_{t}^{t + \tau_{y}} \ud t' \, \int \ud \nu \, T_{00\nu}  \,  e^{i \, \nu \, t'} 
\,  _{\overrightarrow{t \to \infty}}  \\
& _{\overrightarrow{t \to \infty}} \,  \frac{1}{\sqrt{4 \, \pi}} \, \frac{1}{\tau_{y}} \, 
\frac{1}{\sqrt{\tau_{y}}} \, \int_{\tau_{y}} \, \ud t' \,  \sum_{M}  \, T_{00M}  \,  e^{i \, \frac{2 \, \pi}{\tau_{y}} \, M \, t'}  = \frac{1}{\sqrt{4 \, \pi}} \,
\frac{T_{000}}{\sqrt{\tau_{y}}} \\
\end{split}
\end{equation}
This is in agreement with the fact that the solution of a differential system lying on the attractor has a smaller number of dimensions;
in this formulation this statement means that the continuous variable $\nu$ tends to the discrete index $M$.

\end{enumerate}

Dimensional remarks for the expansion in $t$.
\begin{equation*} 
\begin{array}{l}
\displaystyle{\text{dim} \, (T) = \text{K} } \\
\displaystyle{\text{dim} \, (W_{\ell, m}^{M} (\theta, \varphi, t) ) = \text{s}^{-\frac{1}{2}} } \\ 
\displaystyle{\text{dim} \, (W_{\ell, m}^{\nu} (\theta, \varphi, t) ) = \text{dimensionless} } \\ 
\displaystyle{\text{dim} \, (T_{\ell m M}) = \text{K}  \cdot \text{s}^{\frac{1}{2}} } \\ 
\displaystyle{\text{dim} \, (T_{\ell m \nu}) = \text{K}  \cdot \text{s}    } \\ 
\displaystyle{\text{dim} \, (\delta (\nu)) = \text{s}} \\
\end{array}
\end{equation*}  

\newpage

\section*{Appendix D. The numerical calculation of the field $T$}
\renewcommand{\theequation}{D.\arabic{equation}}
\renewcommand{\thefigure}{D.\arabic{figure}}
\setcounter{figure}{0}
\setcounter{equation}{0}
The equation of motion is
\begin{equation} \label{eq:d1}
c_{v} \, \rho_{v} \, \frac{\de T}{\de t} = 
  \kappa \,  \Delta_{r, \theta, \varphi} \, T + \phi^{in} (\theta, \varphi, t) \, \delta (r - R)
\end{equation}
with
\begin{equation*} 
\begin{array}{ll}
- \kappa \, \nabla T \cdot \vec{n} = \sigma \, T^{4}; \qquad  & r = R \\
- \kappa \, \nabla T \cdot \vec{n} = 0; \qquad  & r = R_{1} \\
\end{array}
\end{equation*}
Since the input term is written explicitly in polar coordinates, this imposes for the discrete spatial grid the 
polar choice. $T$ is then the unknown matrix with three indexes, $T_{m,i,j}$ and these indexes are defined as follows:
\begin{equation}  \label{eq:d2}
\begin{array}{lll}
0 \le m \le m_{max}; & R_{1} \le r \le R; & h_{r} = \displaystyle{\frac{R - R_{1}}{m_{max}}}; \\  \\
0 \le i \le i_{max}; & 0 \le \varphi \le 2 \, \pi; & h_{i} = \displaystyle{\frac{2 \, \pi}{i_{max}}}; \\ \\
0 \le j \le j_{max}; &  0 \le \theta \le  \pi; & h_{j} = \displaystyle{\frac{\pi}{j_{max}}}; \\
\end{array}
\end{equation}
The input term in (\ref{eq:d1}) is a given matrix with two indexes:
\begin{equation*} 
\phi^{in} (\theta, \varphi, \psi) \quad \to \quad 
\phi^{in} (i, j, F^{-1} (t) )
\end{equation*}
namely $m$ is fixed to the surface value $m = m_{max}$ and $i, j$ run according to (\ref{eq:d2}).
The operators gradient and Laplacian are written in a standard way as 
relationships between respectively first and second neighbours
\cite{ref:rec}.

The step of the time enters in the derivative $\frac{\ud T_{m,i,j}}{\ud t}$  and $\phi^{in} $.
The time appears in $\phi^{in}$  in two places:
\begin{equation*} 
\begin{array}{l}
\sin \omega_{d} t  \qquad \text{or} \quad \cos \omega_{d} t \\
\sin \psi (t)  \qquad \text{or} \quad \cos \psi (t)  \\
\end{array}
\end{equation*}
The function $\psi (t_{i}) $ must be evaluated numerically at each step $t_{i}$.
\\
The calculation begins with an initial value
\begin{equation*} 
T_{m,i,j} (t = 0) = T_{0}
\end{equation*}
and proceeds with various choices of $\Delta t, \, m_{max}, \, i_{max}, \, j_{max}$. The integration in $t$ is calculated with the Runge Kutta method 
(4th order) \cite{ref:rec}.
The attractor regime is found when the value  $T_{m,i,j} (\nu)$    evaluated after a number  of time steps $\nu = \frac{\tau_{y}}{\Delta t}$
is sufficiently close to $T_{m,i,j} (0)$.
\\
In practice, for our model,  we have used
\begin{equation*}
m_{max} =  20; \qquad i_{max} =  30;  \qquad j_{max} =  40; 
\end{equation*}
\begin{equation*}
 \Delta t = 3 \, \text{minutes};
\end{equation*} 
\begin{equation*}
c_{v} = 840 \, \text{J kg}^{-1} \, \text{K}^{-1}; \qquad   \rho =   2000 \,  \text{kg m}^{-3};
\qquad \kappa  = 1.5 \, \text{W m}^{-1} \, \text{K}^{-1} 
\end{equation*}
For $T_{0} = 200 \, \text{K}$ the transient time required to reach the attractor is $\sim 5 \, \tau_{y} $.
For $R - R_{1}$ we have chosen the value
\begin{equation*}
 R - R_{1} \sim 6 \, \text{m}
\end{equation*} 
This value is approximatly twice the penetration lenght $\tilde{\delta}$
of the mode $M = 1$ corresponding to one year discussed in appendix C.
This is the general plan of the numerical study. Of course
for the study of the daily cycle the penetration lenght is smaller and the steps in $r$ are accordingly concetrated near $r < R$.

\end{document}